# Scalable additive manufacturing via integral image formation

Seok Kim[1], Jordan Jerome Handler[2], Young Tae Cho[3], George Barbastathis[1,4], and Nicholas Xuanlai Fang[1]*

[1]Department of Mechanical Engineering, Massachusetts Institute of Technology, Cambridge, MA 02139, USA.
[2]Sloan School of Management, Massachusetts Institute of Technology, Cambridge, MA 02142 USA.
[3]Department of Mechanical Engineering, Changwon National University, Changwon, South Korea.
[4]Singapore-MIT Alliance for Research and Technology (SMART) Centre, 1 Create Way, Singapore 138602.
*e-mail: nicfang@mit.edu

## Abstract

Additive manufacturing techniques enable the fabrication of functional microstructures with mechanical and chemical properties tailored to their intended use. One common technique is stereolithography, which has recently been augmented in response to modern demands to support smaller features, larger build areas, and/or faster speeds. However, a limitation is that such systems typically utilize a single-aperture imaging configuration, which restricts their ability to produce microstructures at large volumes due to the tradeoff between the image resolution and image field area. In this paper, we demonstrate three versatile imaging functions based on the coupling a planar lens array, namely, parallel image transfer, kaleidoscopic superposition, and integral reconstruction, the combination of which is termed integral lithography. In this approach, the individual microlenses in the planar lens array maintain a high numerical aperture and are employed in the creation of digital light patterns that can expand the printable area by the number of microlenses ($10^3$-$10^4$). The proposed lens array-based integral imaging system provides the concurrent ability to scale-up and scale-down incoming image fields, thereby enabling the scalable stereolithographic fabrication of three-dimensional features that surpass the resolution-to-area scaling limit. The proposed system opens up new possibilities for producing periodic microarchitectures spanning four orders of magnitude from micrometers to centimeters that can be applied to biological scaffolds, metastructures, chemical reactors, and functional surfaces.

## Introduction

Rapid developments in the fabrication of three-dimensional (3D) printed architectures has revolutionized the ability to produce functional structures for mechanical/acoustic metamaterials [1-3], cellular mechanobiological materials [4], and structures for energy/environmental applications [5, 6]. For instance, 3D microstructures with mechanically compliant materials and customized constructed scaffolds offer tailored functionality that satisfies the criteria for biocompatibility and defined stiffness [4]. Moreover, the application of functional structures in catalytic system have improved efficiencies by way of micro- and nanoscale architectures designed to increase surface area-to-volume ratios with reduced mass [5, 6]. Also, advances in additive manufacturing techniques have enabled the fabrication of these functional structures with complex architectures on various spatial scales down to the sub-micrometer scale [7-10]. The commonly used stereolithography (SLA) technique supports the fabrication of high resolution and geometrically complex manufactured products [7, 8], and recent advances have significantly improved on the feature resolution [11, 12], speed [13], and build size [14-17]. For instance, digital micro-mirror devices [11] and spatial light modulators [14] has enabled the curing of large areas (termed projection micro-stereolithography (PμSL)), as opposed to the traditional 'tracing' approach employed by single or multiple spot laser systems [18]. Recent works have demonstrated variants of PμSL that incorporate a serial printing process in which many repeated scanning cycles expand the overall build size without sacrificing resolution [14-17]. One recent derivative of PμSL, termed volumetric printing, overcomes the current layer-by-layer manufacturing approach to almost instantaneously fabricate 3D objects [19, 20].

However, despite these system improvements, conventional PμSL methods use an imaging platform that relies on a single-aperture imaging system in which an incoming image is focused directly onto a single planar area. The consequence is that the amount of transferred spatial information is fundamentally limited by the space-bandwidth product (SBP) of the pixelated digital projection system, which is defined as the number of pixels required to realize the maximum information capacity. The SBP of a conventional PμSL platform is typically in megapixels (Mpx) range regardless of the numerical aperture (***NA***) or magnification (***M***) of the imaging optics used, which results in a compromise between the achievable feature resolution and the total image area [8, 21]. This compromise must be overcome to further advance microstructural 3D printing into production use.

This problem can potentially be solved with an image multiplication strategy (i.e., numbering-up) in conjunction with a planar micro-optical imaging system. With continued advances in low-cost and large scale microlens array fabrication techniques, micro-optical devices have become a promising tools for large-area display applications such as integral imaging 3D displays [22]. A benefit of these fabrication techniques is that they have already been proven scalable over the past few decades. Image multiplication via micro-optical imaging devices has been demonstrated in Talbot array illumination [23] and microlens projection lithography [24, 25], which are capable of fabricating sub-micrometer two-dimensional (2D) lattice structures. However, the use of a static photomask limits the imaging function to a simple duplication of a single object, and therefore does not satisfy the design requirements for complex architectures with multiple layers beyond 2D planar structures.

At present, both micro-optical and single-aperture imaging systems require further development and no existing technologies can support a scalable SBP in additive manufacturing. Here, we propose a new stereolithographic printing system that combines integral image formation with a planar micro-optical device to provide a scalable SBP for additive manufacturing applications without requiring serial scanning. The proposed engineered projection system is based on a lens array, in which each individual microlens can maintain a high ***NA*** and the overall print area can be expanded by the number of microlenses. The functional micro-optical device combined with digital light processing enables the projected output images to be manipulated via versatile projection functions of the incoming images, namely, parallel transfer, kaleidoscopic superposition, and integral projection. We developed an approach, integral lithography, to realize the concurrent ability to scale-up (increasing areal build size without compromising the critical feature size) and scale-down (reducing the feature size while maintaining the areal build size) incoming image fields. The imaging mechanism is described with a simple thin-lens equation that predicts the complex output patterns of the lens array from simple inputs. Furthermore, we demonstrate that the reconstructed patterns obtained from the integral imaging of multiple incoming images allows us to improve the scalability of the print area and provide it with a more uniform light distribution. As part of our validation, we evaluated the scalability of the approach and its ability to expand print areas $10^2$-$10^3$ times larger than those of current commercial systems, which translates to an SBP of 0.1-0.28 gigapixels (Gpx). Of particular interest was the ability to produce periodic microarchitectures spanning four orders of magnitude in scale from micrometers to centimeters.

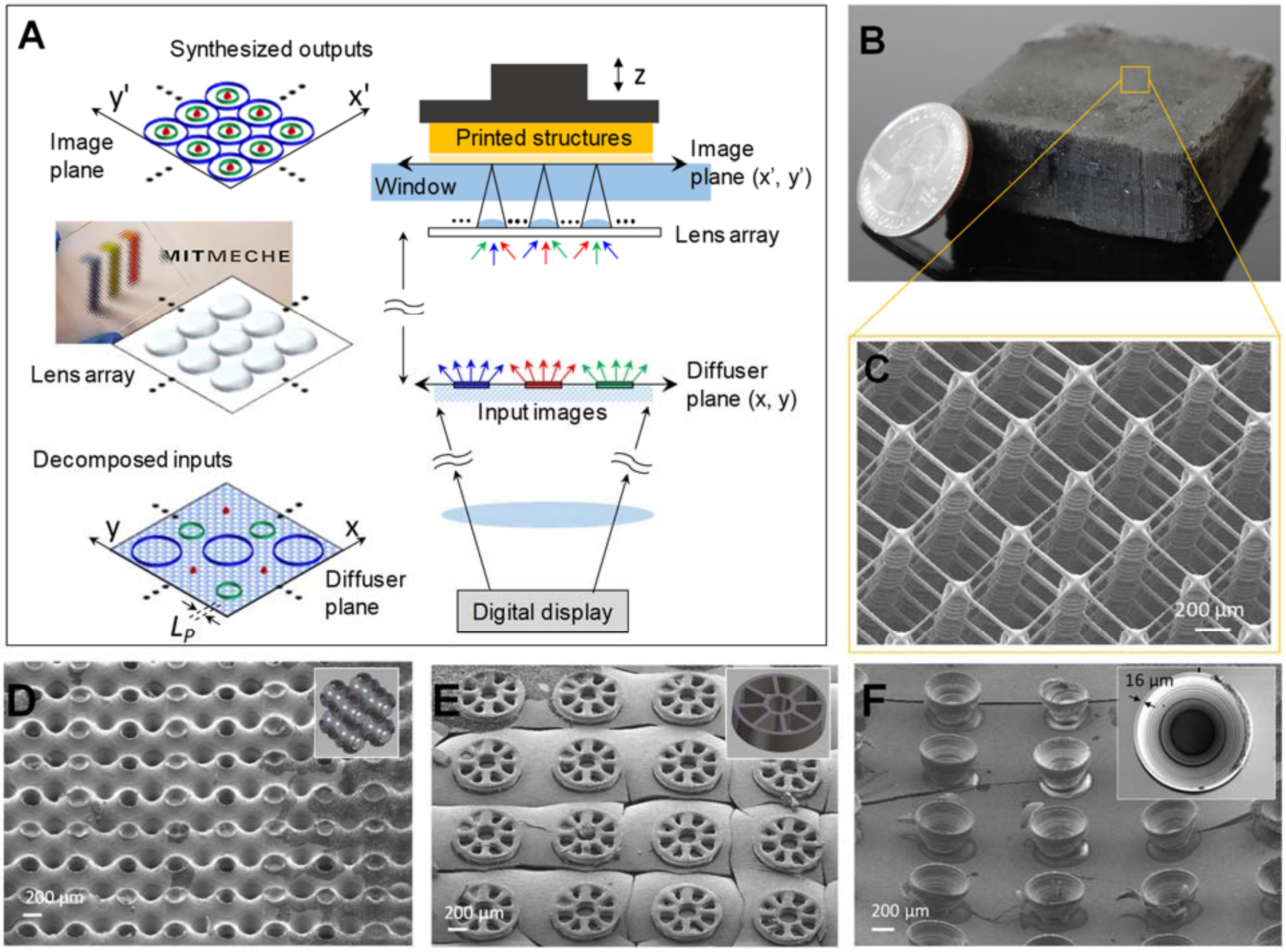

**Fig. 1: Integral lithographic system for scalable additive manufacturing.** (A) Schematic view of the integral lithographic system. The reconstructed imaging patterns are projected by the lens array (displayed in front of the MIT mechanical logo) in conjuction with the digital microdisplay. (B to F) Microstructures were fabricated layer-by-layer with exposure time 3-30 seconds under an intensity of 33 mW/cm$^2$. These multiscale structures were produced by a lens array (Lens 1, defined within the caption of Fig. 4 and in the Methods section): (B) cubic-truss microlattices (total layers with a slicing thickness of 5-50 μm: 400), and (C) scanning electron micrograph of microlattices with strut suspended beam diameter of 5 μm; (D) triply

periodic bicontinuous structures (total layers with a slicing thickness of 20 μm: 60); (E) circular-lattice microscaffolds (total layers with a slicing thickness of 10 μm: 10); (F) trapezoidal shell-type microstructures with a reentrant geometry (total layers with a slicing thickness of 20 μm: 20).

**Engineered imaging system**

A schematic overview of the proposed 3D printing system is shown in Figure 1A. As shown, a digitally generated object image is projected onto a diffuser, which acts as the input image plane in our system, and is observed by the lens array [25-28]. The lens array (displayed in Fig. 1A in front of the MIT mechanical logo) is capable of focusing light sources from multiple viewpoints to both replicate and reconstruct images into altogether new patterns [29-31], and this functionality allows the the incoming images to be superimposed and integrally reconstructed. By leveraging the versatile imaging functions in conjunction with the lens array with the microdisplay device, the engineered projection-based printing system enables the high-resolution and scalable stereolithographic manufacturing of complex microstructures. During the printing process, multiple output images, each generated by a unit-lens of the lens array, form reconfigurable synthetic patterns via one or more combinations of parallel replication, kaleidoscopic superposition, and integral reconstruction. A set of these reconstructed images is then used to create 3D architectures via PuSL techniques in which the photopolymer is cured in a layer-by-layer fashion. The prints of the complex 3D microstructures shown ing Fig. 1B-F and S1 contain minimum feature sizes of ~ 5-20 μm over an area several tens of $cm^2$ and demonstrate the feasibility of printing polymeric structures that exceed the resolution-to-area scaling limit. The microlattices in Fig. 1B-C were fabricated with 400 layers of reconstructed output patterns with a layer thickness of 5-50 μm. The cubic-truss lattice in these figures is composed of three freestanding mesh layers suspended on an array of vertical posts and separated by identical distances in the vertical direction. These polymeric microlattices can be utilized at scale in tailored mechanical environments, such as to mimic artificial axons [4] or form a catalytic reactor with a high surface area-to-volume ratio [33]. Our approach enables the fabrication of complex 3D microstructures that are difficult to achieve with conventional projection lithography processes. For example, by varying the geometric overlap of the image outputs from each unit-lens, we 3D-printed a wide variety of structures from interconnected bicontinuous structures (Fig. 1D) to isolated microarchitectures of circular-lattice scaffolds (Fig. 1E) and trapezoidal reentrant structures (Fig. 1F). Note that these examples of 3D periodic microstructures with different degrees of connectivity can be extended to a variety of tissue scaffolds [34], mechanical metamaterials [35], feed spacers for water reuse system [36], or functional surfaces [37].

A description of how the imaging mechanism manipulates the projected output patterns is provided in Figure 2 and the geometric relationship between the lens array, input image, and output image is shown in Fig. 2A, where it can be seen that the input image information is transferred in parallel by the lens array to generate an array of repetitive patterns that can produce complex patterns beyond simple replicated images. The relationship between the input image size ***a***, the output image size ***a'***, and overlap of resulting output images is as per the simple thin lens equation of $1/\boldsymbol{f} = 1/\boldsymbol{b} + 1/\boldsymbol{b'}$ [32], where $\boldsymbol{f}$ is the effective focal length of the unit-lens, $\boldsymbol{b}$ is the distance between the lens array and the input image plane, and $\boldsymbol{b'}$ is the distance from lens array to the output image plane. The demagnification factor of the output image from each unit-lens is defined as $\boldsymbol{D} = \boldsymbol{b'}/\boldsymbol{b} = \boldsymbol{f}/(\boldsymbol{b} - \boldsymbol{f})$ and the resulting $\boldsymbol{a'}$ of each lens unit is $\boldsymbol{a} \times \boldsymbol{D}$ as shown in Fig. 2A. The lens array is mounted on a micro-translation stage, which allows to longitudinal movement along the $z$-axis to control $\boldsymbol{D}$ of the output image by adjusting $\boldsymbol{b}$ from the input image plane. Note, we assume that the unit-lens size is equal to the lattice spacing $\boldsymbol{p}$ of the lens array. When $\boldsymbol{a'}$ is larger than $\boldsymbol{p}$, the multiplied images interconnect and overlap with each other to reproduce kaleidoscopic patterns in the same imaging plane. Unlike the illumination based approach in prior work [31], the kaleidoscopic approach in this work stems from the superposition of multiple output images and enables more controllable and predictable patterns through combinations of the dynamic input image and lens array imaging. Figures 2B and S2 shows representative output patterns and printings created by modulating the geometry and size of the input images via the lens array.

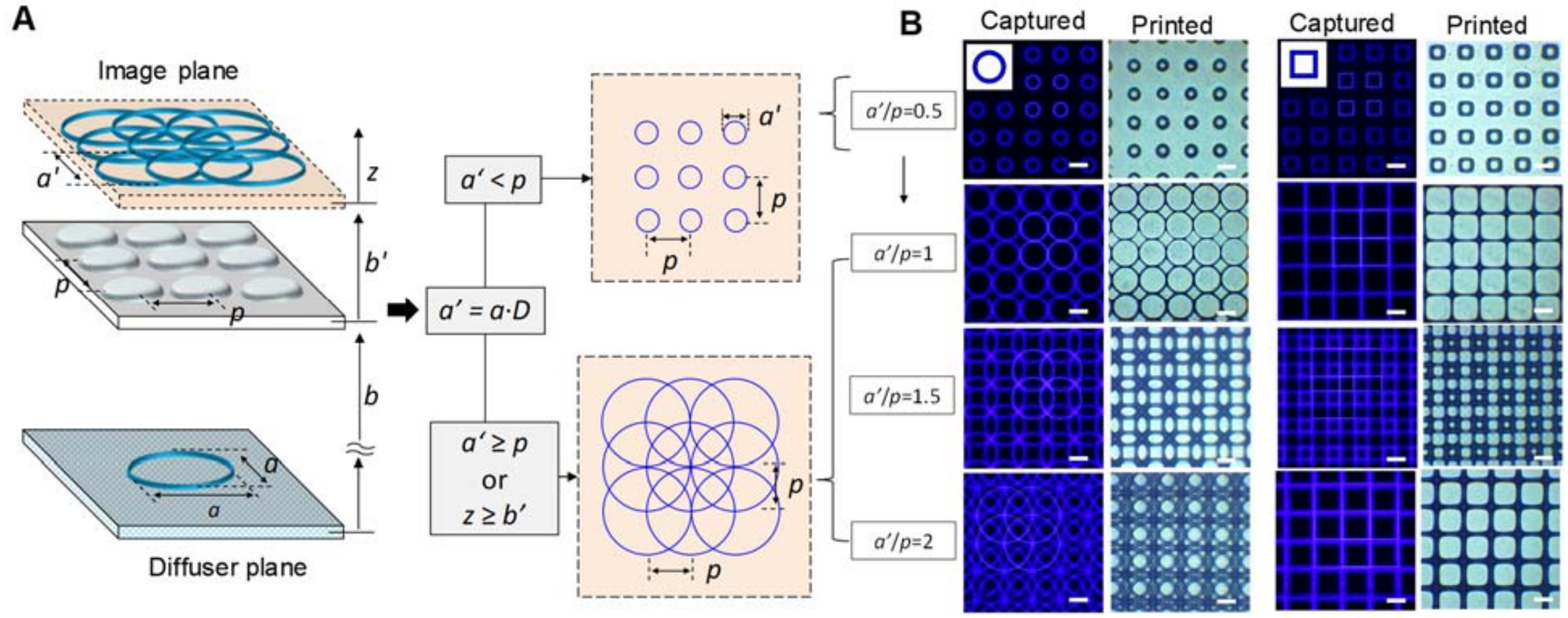

**Fig. 2: Digitally controlled imaging patterns.** The geometric relationship between the lens array and an input object produces kaleidoscopic patterns. The synthetic images are projected on an imaging plane of the lens array, captured by an optical microscope, and recorded within the photopolymer resin via photopolymerization. (A) Parallel replication of a single object image by the lens array, which captures an object image and generates an array of repetitive patterns. Kaleidoscopic patterns form through the overlap and superposition of multiple replicated images based on the interaction between the lens array and a single object image. (B) Optical microscope-captured images and corresponding polymerized microstructures for various kaleidoscopic structures. When $\boldsymbol{a'}/\boldsymbol{p} < 1$, parallelized isolated images are produced onto an imaging plane, whereas when $\boldsymbol{a'}/\boldsymbol{p} \geq 1$, interwoven patterns are reproduced by connecting and overlapping the replicated patterns. Note that this approach can be used to realize out-of-plane kaleidoscopic patterns with periodic optical distributions at different propagation distances ($\boldsymbol{z} \geq \boldsymbol{b'}$) (Fig. S3). All scale bars are 100 µm.

The homogeneous light distribution into the lens array from the diffuser enables images from different perspectives (i.e., not orthogonally projected) to be combined in reconstruction pcoress [25, 27, 28]. In contrast to patterns generated patterns based on the parallel transfer and superposition of a single input image, as shown in Fig. 2, these synthetic patterns stem from imaging techniques analogous to the integral imaging techniques used in a multiview 3D display [26]. Each unit-lens of the lens array can observe multiple elemental images (EIs) and reconstructing them into identical and/or highly periodic composited patterns, as shown in Fig. 3.

To describe the relationship between the elements of the input objects and output images, we consider the optical system of a one-dimensional model with the column vectors $\boldsymbol{a_{in}^n}$, $\boldsymbol{a_{out}}$, and matrix $\boldsymbol{H}$, where $\boldsymbol{a_{in}^n}$ and $\boldsymbol{a_{out}}$ are the elements of the input objects and the projected images, respectively, and $\boldsymbol{H}$ denotes the optical system matrix (see the Methods section). Then, the system can be described as $[\boldsymbol{a_{out}}] = [\boldsymbol{H}][\boldsymbol{a_{in}^n}]$, as shown in Fig. 3A. The spacing $\boldsymbol{A_{in}}$ of the EIs is also reduced by a factor of $\boldsymbol{D}$ and forms the spacing $\boldsymbol{A_{out}}$ of the output image array. A geometrical condition described as $\boldsymbol{A_{out}} = \boldsymbol{p}$ and $\boldsymbol{a_{out}} \leq \boldsymbol{p}$ enables the multi-view reconstruction by the superimposed images, which allows multiple sub-images to be used in the construction of a desired composited pattern or to create a continuous networked pattern. This allows the creation of scalable projected patterns in stereolithographic additive manufacturing. To prove the concept of the integral imaging patterns, Figure 3B illustrates the synthesized imagery created by digitally interlacing a set of EIs with indentical ($\boldsymbol{a_{in}^1} = \boldsymbol{a_{in}^2} = \boldsymbol{a_{in}^3}$) or three decomposed ($\boldsymbol{a_{in}^1} \neq \boldsymbol{a_{in}^2} \neq \boldsymbol{a_{in}^3}$) spatial components, respectively. In both cases, the input objects were spatially multiplexed and decoded as the synthetic images via integration in the imaging plane of the lens array. Since the illumination sources in this case are incoherent, the intensity distribution of the synthetic images from the lens array can be assumed to be a simple linear superposition of all reduced EIs. The overall surface topologies and cross-sectional intensity profiles of the projected patterns (Fig. 3B) confirm the consistency between the composite patterns created via integral imaging (see Fig. S4-5 in Supplementary Information).

Notably, the integral imaging with the sparse spacing of decomposed EIs, termed the compressive integral imaging in this study, can provide considerable benefits when coupled with a cheap, low bandwidth display units (e.g., sub-Mpx). According to a simple one-dimensional model assumption (Fig. S6), the pixel values of the display image $\boldsymbol{a_{in}^n(x)}$ can be represented by $f^n(\boldsymbol{x})\text{comb}(\boldsymbol{x}/\boldsymbol{L_P}) * \text{rect}(\boldsymbol{x}/\boldsymbol{L_P})$, where $f^n(\boldsymbol{x})$ and $\boldsymbol{L_P}$ represent the target object and the projected pixel size of the constituent digital display, respectively. The Fourier transform of $\boldsymbol{a_{in}^n(x)}$ is represented by $\boldsymbol{A_{in}^n}(\boldsymbol{v_x}) = |\boldsymbol{L_P}^2| F^n(\boldsymbol{v_x}) * \text{comb}(\boldsymbol{v_x}\boldsymbol{L_P})\text{sinc}(\boldsymbol{v_x}\boldsymbol{L_P})$, where $\boldsymbol{v_x}$ is the spatial frequency (= $1/\boldsymbol{C}$) of the target image. Considering the Nyquist sampling criteria ($\boldsymbol{v_N} = 1/2\boldsymbol{L_P}$), a digital micromirror device (DMD)-based display unit ($\boldsymbol{L_P}$ of ~ 50 μm) can provide enough spatial resolution ($|\boldsymbol{v_x}| < \boldsymbol{v_N}$) to prevent the aliasing (i.e., spectral overlap) in integral imaging with both identical and decomposed EIs, thereby enables comparable reconstruction performance, as shown in Fig. 3B. However, a microdisplay with $\boldsymbol{L_P}$ of ~ 220 μm results in the aliasing from the integral imaging of identical EIs (Fig. 3C-i) due to insufficient spatial resolution ($|\boldsymbol{v_x}| > \boldsymbol{v_N}$). Herein, the compressive integral projection used to decomposed the high-frequency spatial component of the initial target image (i.e., $|\boldsymbol{v_x}| < \boldsymbol{v_N}$) can provide a solution that can restore the desired target image (see Fig. S6 in Supplementary Information). As shonw in Figure 3D-ii, the integration of micro-optical elements into realatively cheal display devices can resolve high frequency components without aliasing effects, and is expected to facilitate low-cost mass-production. Although the integral projection of only three decomposed EIs was used to create the desired target image in this study, we expect that further optimal solutions may be obtained by addressing the inverse problem that arises when low bandwidth sub-images are used to synthesize high bandwidth images.

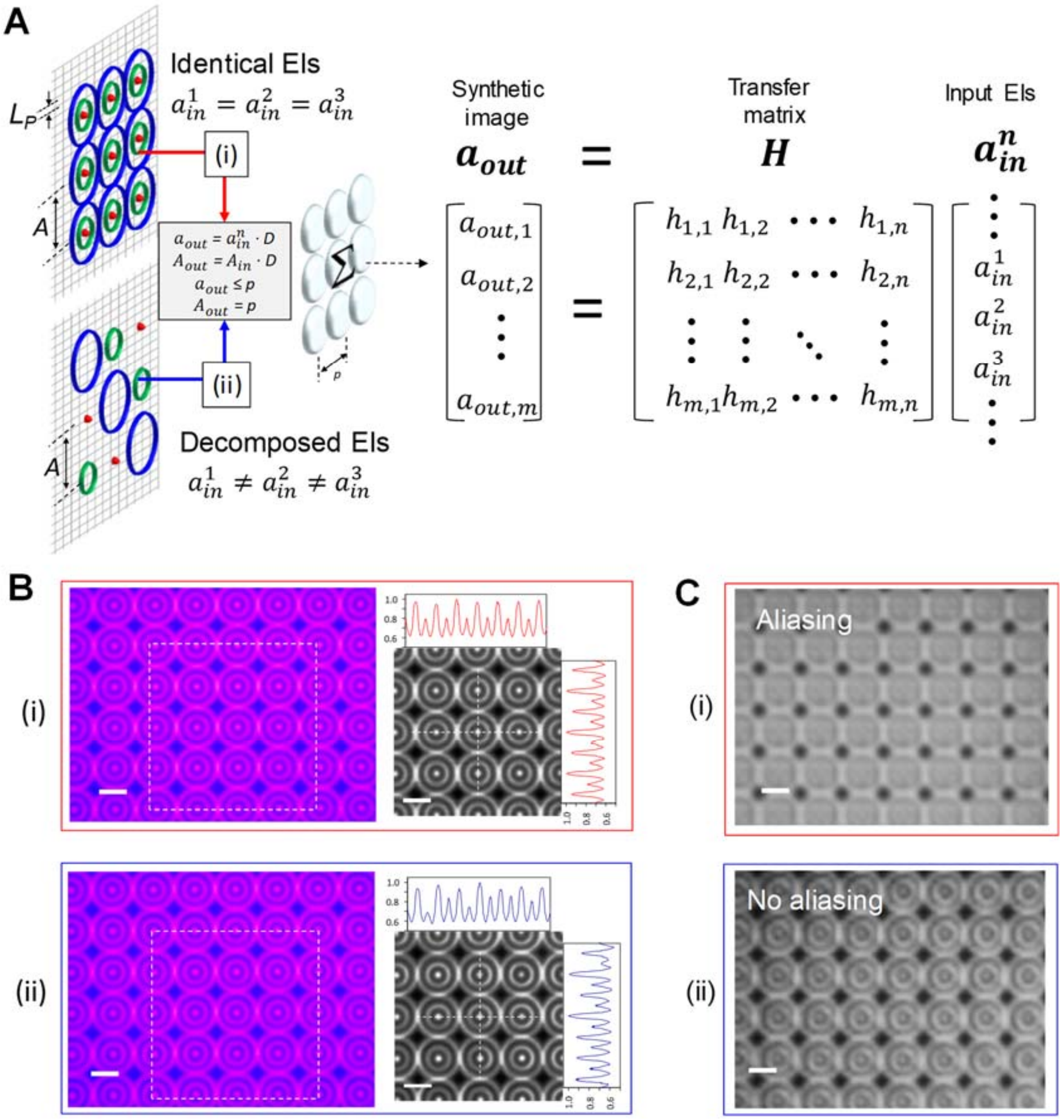

**Fig. 3: Integral imaging patterns with compressive multi-projection.** (A) Geometrical relationship and matrix form of the integral image formation between input objects of identical/decomposed EIs and projected outputs. The transfer matrix $\boldsymbol{H}$ is determined by its elements $\boldsymbol{h_{m,n}}$, which represent the impulse response function of the unit-lens in the lens array, and $\boldsymbol{m}$ and $\boldsymbol{n}$ represent the numbers of the unit-lenses within the lens array in the horizontal direction and vertical direction, respectively. (B) Optical microscope-captured topologies and cross-sectional intensity profiles of integral imaging patterns created by a digital

microdisplay with an $\boldsymbol{L_P}$ of 50 µm. The intensity profiles were normalized to the maximum gray value versus the pixel distance. (C) Output imaging patterns created by a microdisplay with an $\boldsymbol{L_P}$ of 220 µm. (i) Integral imaging with indentical EIs of a concentric circular grating. (ii) Integral imaging with the three decomposed EIs for synthesis into the concentric circular grating on the imaging plane through the lens array. All scale bars are 100 µm.

**Scalable photopolymerization**

The coupling of digitally-controlled integral imaging patterns with a lens array enabled the scalable microprinting of various structures. Intertwined fibrous lattice microstructures were printed under Lens 1 (defined within the Fig. 4 caption and the Methods section) with a minimum feature size of ~ 5 µm over an exposure area up to 2500 mm$^2$ (Figs. 4A-C and S1E-H). Arbitrary patterns comprised of both array lines (Fig. 4F-K) with feature sizes down to 1-2 µm and array letters of 'MiT' with 50 µm length were fabricated with Lens 2 (defined within the Fig. 4 caption and the Methods section). Considering a several mm$^2$ exposure area and a similar lateral feature size of the single-aperture imaging based PµSL configuration [11, 12], the areal ratio (~10$^2$) of printing scales demonstrates that this imaging approach as scalable without compromising optical resolution. Furthermore, the imaging approach provides new opportunities in applications requiring the high-throughput fabrication of custom-shaped microparticles or micro-textured surfaces. For example, flexible multiarm particles (Fig. 4D), micro-wavy patterned surfaces (Fig. 4E), or 3D microparticles with a microwell arrays (Figs. 4J-K) can be fabricated to serve as customized microstructural platforms for efficient cell-capture in the detection and characterization of circulating cells [37-39].

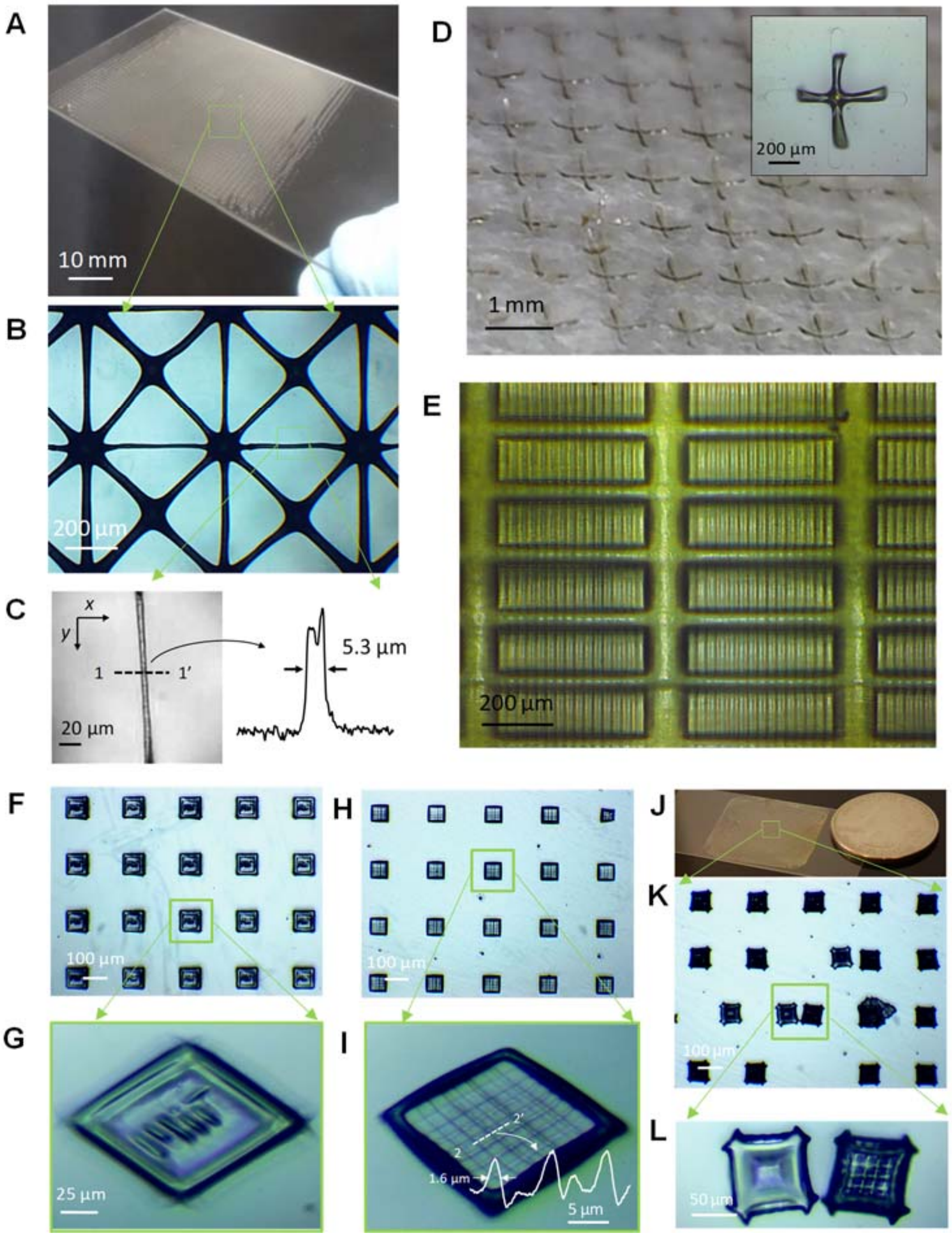

**Fig. 4: Scalable printing with small feature sizes.** (A-E) Micro-structures/particles created with Lens 1 (***f*** = 5.5 mm, ***p*** = 1 mm, effective ***NA*** of 0.14 in the photopolymer, and an overall size of 50 × 50 $mm^2$): (A-C) periodic microstructures, such as fibrous lattice, with a minimum feature size of ~ 5.3 µm over an exposure area up to 2500 $mm^2$; (D) flexible multiarm microparticles; and (E) micro-textured surfaces. (F-K) Arbitrary micro-patterns/particles fabricated with Lens 2 (***f*** = 0.57 mm, ***p*** = 0.25 mm, effective ***NA*** 0.33 in photopolymer, and an overall size 25 × 25 $mm^2$): (F-I) Array lines with feature sizes down to ~ 1.6 µm and array letters 'MiT' with a maximum exposure area up to 625 $mm^2$; and (J-K) 3D microparticles with a microwell array. All microstructures were printed by integral imaging patterns of identical EIs with a single exposure of 3 to 10 s under an intensity of 33 $mW/cm^2$. The line profiles of the optical images in (C) and (I) were quantitatively analyzed using the imaging software package ImageJ.

## Discussions

In single-aperture imaging systems based on a pixelated digital microdisplay [40], an areal build size ($\boldsymbol{A_S}$) in mm$^2$ during unit exposure is defined as $(total\ pixels) \cdot (\boldsymbol{L_D}/\boldsymbol{M})^2$, where $\boldsymbol{L_D}$ is the display pixel size of the constituent digital microdisplay and $\boldsymbol{L_D}/\boldsymbol{M}$ is equal to $\boldsymbol{L_P}$. A rational strategy for improving the resolution is to increase $\boldsymbol{M}$ to reduce of $\boldsymbol{L_D}$. However, reducing $\boldsymbol{L_D}$ by a × 10 magnification lens (i.e., $\boldsymbol{M}$ = 10) will result in a 100 times reduction of the $\boldsymbol{A_S}$. Thus, the scaling problem of increasing the $\boldsymbol{A_S}$ without compromising resolution remains a challenge in PμSL. To investigate the effect of the integral lithographic system on the scaling issue, we analyzed the $\boldsymbol{A_S}$ and minimum feature size ($\boldsymbol{R}$) for a range of existing PμSL products with available digital microdisplay devices. On the $\boldsymbol{A_s}$ - $\boldsymbol{R}$ plot shown in Fig. 5, $\boldsymbol{R}$ is rendered as $(\boldsymbol{L_D}/\boldsymbol{M})$ [41, 42]. The scaling limit, which is the ability of existing projection-based 3D printing technologies to scale microstructures, is shown in Figure 5. The empirical scaling behavior shown was deduced from the specific published specifications of PμSL machines (gray square dots in Fig. 5). Following the apparent scaling dependence of the PμSL approach, a theoretical analysis predicts the relationship $\boldsymbol{A_S} = \boldsymbol{k} \cdot \boldsymbol{R}^2$ [41, 42], where $\boldsymbol{k}$ is the scaling constant corresponding to the total pixels within available digital microdisplay devices [42-45] and refers to the SBP (i.e., the total amount of the transferable information capacity) in the optical imaging system. In Fig. 5, these analytic scaling boundaries are denoted with dashed lines, where the red and green circle-dots represent the experimental results and expected potential, respectively, of the proposed printing system using the current optical system, respectively. The areal build size ($\boldsymbol{A_I}$) of integral lithography exhibits different scaling constants from the $\boldsymbol{A_S}$ - $\boldsymbol{R}$ scaling of other PμSL versus the resolution and minimum feature size. Based on the empirical illumination distribution in our system, the achievable maximum condition could be described as $\boldsymbol{A_I} \leq \boldsymbol{A_S}$ because the uniform illumination region and its resulting $\boldsymbol{A_S}$ are determined by the maximum area of the virtual imaging mask to be observed by the lens array [46]. Considering this condition, we estimated the $\boldsymbol{A_S}$ - $\boldsymbol{R}$ relationship for the integral lithography method to compare its performance with that of current PμSL analyzed in Fig. 5. The effective planar resolution $\boldsymbol{R_{eff}}$ of the lens array is assumed be described by $\boldsymbol{R} \times \boldsymbol{D}$ by considering the geometric optics. The corresponding equation can be expressed as:

$$\boldsymbol{A_I} \leq \boldsymbol{A_S} = \boldsymbol{k} \cdot \boldsymbol{R}^2 = \boldsymbol{k} \cdot \left(\frac{\boldsymbol{R_{eff}}}{\boldsymbol{D}}\right)^2 = \left(\frac{\boldsymbol{k}}{\boldsymbol{D}^2}\right) \cdot \boldsymbol{R_{eff}}^2 = \boldsymbol{k_{eff}} \cdot \boldsymbol{R_{eff}}^2, \quad (1)$$

where $\boldsymbol{k_{eff}}$ is $\boldsymbol{k}/\boldsymbol{D}^2$ and $\boldsymbol{R_{eff}}$ must be compliant with the Abbe diffraction-limited spot size $\boldsymbol{d} = 1.22\boldsymbol{\lambda} / 2\boldsymbol{NA}$ [47], where the $\boldsymbol{NA}$ of the unit-lens is defined by $\boldsymbol{n}\sin(\tan^{-1}(\boldsymbol{p} / 2\boldsymbol{f}))$. Here, we assume that the refractive index $\boldsymbol{n}$ of the photopolymers is 1.5. In this work, all printing experiments were performed at an imaging distance $\boldsymbol{b}$ of 68.75 mm and the demagnification factors $\boldsymbol{D}$ for Lens 1 and Lens 2 were 0.087 and 0.0084, respectively, after considering the geometric condition of the lens array. The ideal $\boldsymbol{k_{eff}}$ was calculated as $1.35 \times 10^8$ (~ 0.14 Gpx) and $4.67 \times 10^9$ (4.67 Gpx) for Lens 1 and Lens 2, respectively (the details are provided in the Methods section). From the printed results in this study, we also obtained an experimental $\boldsymbol{k_{eff}}$ of $1 \times 10^8$ (0.1 Gpx) and $2.77 \times 10^8$ (~ 0. 28 Gpx) for Lens1 and Lens2, respectively. As marked at the upper-left side of the lines representing the theoretical scaling plot in Fig. 5, our approach demonstrates the potential to overcome the conventional scaling behaviors of $\boldsymbol{A_S}$ - $\boldsymbol{R}$ (SBP-$\boldsymbol{R}$ plot is shown in Fig. S7). However, we believe that the discrepancy in $\boldsymbol{k_{eff}}$ between the calculated and experimental results does not imply a fundamental limit in our system performance. This is because the limit of $\boldsymbol{A_I}$ depends on the available size of the lens array and digital microdisplay devices, and the obtainable minimum feature size $\boldsymbol{R_{eff}}$ is determined by the overall contribution from the photopolymerization kinetics [11] and the performance of the imaging system.

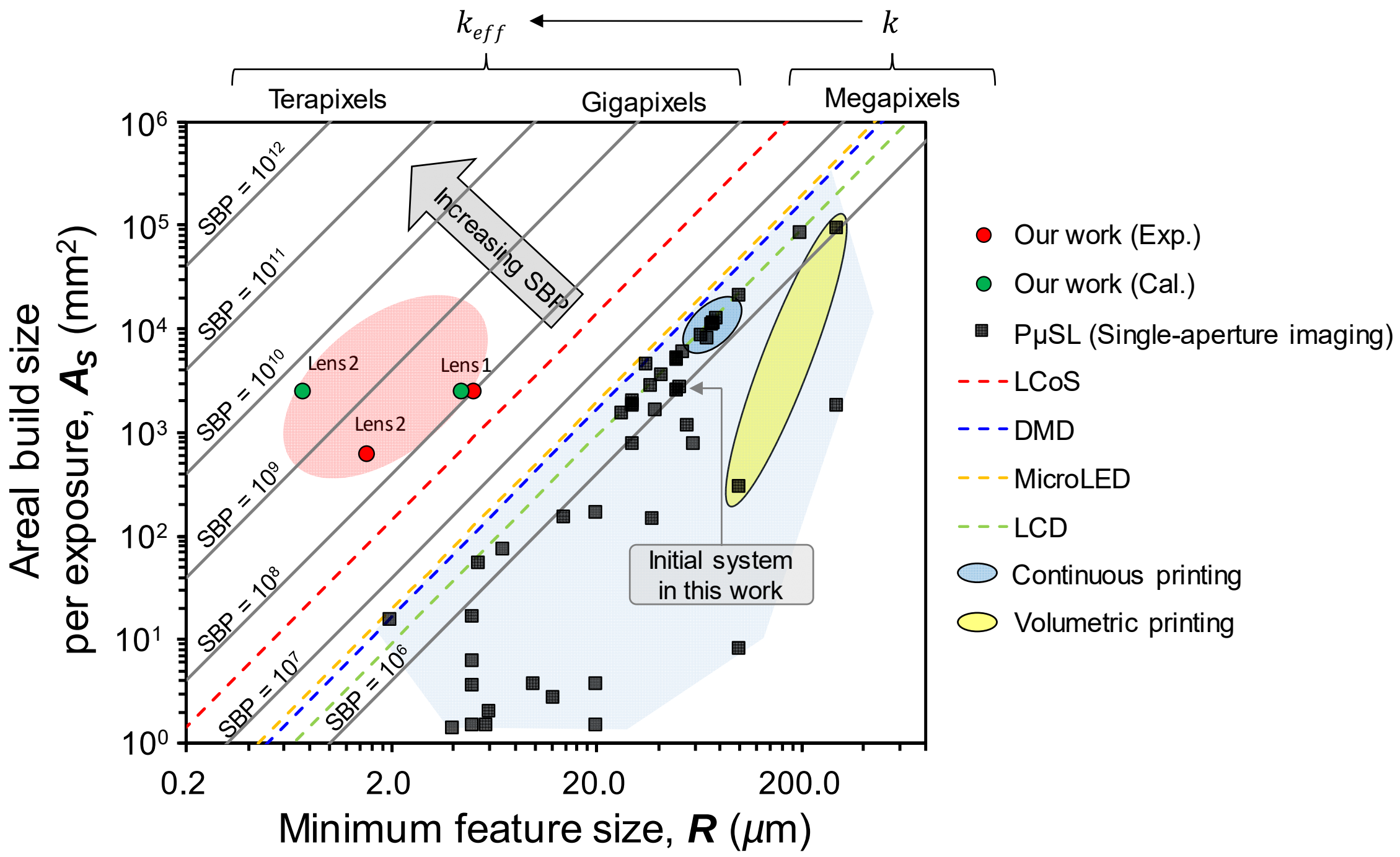

**Fig. 5: Figure of merit for the integral lithographic system.** Comparison of the PμSL methods as a function of the areal build size ($\boldsymbol{A_S}$) versus the achievable minimum feature size ($\boldsymbol{R}$). The dashed line represents analytical scaling equations grouped by digital microdisplay devices of liquid crystal on silicon (LCoS) [42], DMD [43], MicroLED [44], or liquid crystal display (LCD) [45]. The plotted data points from specific published results of PμSL (based on a single-aperture imaging system) exhibit an empirical scaling dependency. The more comprehensive data set used to produce the plot is given Table 1 and Fig. S7 in the Supplementary Information. The red and green circles represent the experimental results as well as the calculations completed by the authors to determine the ideal results representing the potential of integral lithography, respectively. Continuous printing (13); Volumetric printing (19, 20).

We have demonstrated a means of integrating a micro-optical imaging device into a light-based 3D printing system to enable scalable stereolithographic manufacturing via integral image formation. The proposed approach leverages the periodic nature of the geometries produced by reconstructing input images using versatile projection functions consisting of parallel replication, kaleidoscopic superposition, and integral reconstruction based on the complex local geometries generated by the unit-lens. We envision the proposed approach will be used to expand the capabilities of large-area fabrication towards complex and periodic microstructures with mechanical and structural benefits that have yet to be fully utilized at scales that are practical in volume production applications. If such microarchitectures are made accessible at larger scales than currently exist, architected materials like those described in this paper could find widespread applications ranging in biomedical devices [4, 34, 39], extraordinary mechanical systems [35], functional textured surfaces [37], and substrates for energy conversion systems [33, 36, 48]. Moreover, our integral lithographic system could be incorporated into other digital light processing-based lithography with different types and sizes of display systems to enable further increases in the systems' build areas with simple and inexpensive components. This compatibility may motivate the integration of our approach with digital optofluidic fabrication for high-throughput microparticle synthesis [49], or maskless lithography [47]. In summary, our work not only provides a new stereolithographic microfabrication method and platform for scalable 3D printing, it also opens up new possibilities for the mass-production of micro-structures/particles using the versatile imaging functions of planar micro-optical devices.

**Methods**

**Imaging:** The input images in Fig. 2B and 3B were created on a diffuser through the digital displaying of a conventional PμSL system using a DMD-based digital optical engine (Wintech PRO4500) with an $\boldsymbol{L_D}$ of 7.6 μm, $\boldsymbol{M}$ of ~ 1/6.5, and an $\boldsymbol{A_S}$ of 2.56 × $10^3$ $mm^2$. The output images created by the lens array were recorded by using a microscope digital CMOS sensor (AmScope MU500, sensor pixel width of 2.2 μm) with a 2× reduction lens. For the imaging purposes, we placed a lens array of $\boldsymbol{p}$ = 0.15 mm, $\boldsymbol{f}$ = 5.2 mm, and effective $\boldsymbol{NA}$ 0.014 in air (Thorlabs, MLA150-5C, overall size 10 × 10 $mm^2$) was placed at an imaging distance $\boldsymbol{b}$ = 68.75 mm from the masking plane in our system. First, the kaleidoscopic images in Fig. 2 were produced by adjusting the projection image shapes and sizes from 0.92 to 3.66 mm with a $\boldsymbol{D}$ of 0.082. Here, the focal plane of the ditigal microscope camera coincided with the imaging plane of the lens array ($\boldsymbol{z} = \boldsymbol{b'}$). Second, the vertical set position of the ditigal microscope camera was changed until the lens array produced out-of-plane kaleidoscopic patterns. Then, the distance between the focal plane of the microscope and the imaging plane of lens array was gradually increased. The images in Fig. S2-3 were captured as the distance varied ($\boldsymbol{z} \geq \boldsymbol{b'}$). We then arranged EIs of 9 × 9 with distance $\boldsymbol{A}$ of 1.83 mm showing identical or decomposed images of the concentric circular grating to obtain the characterization of the projected patterns in Fig. 3. The input images in the case of Fig. 3C were generated via an LCD-based 3D printing system (Photocentric 3D Liquid Crystal LC10) with an $\boldsymbol{L_D}$ of ~ 220 μm, $\boldsymbol{M}$ of 1, and $\boldsymbol{A_S}$ of 2 × $10^4$ $mm^2$. In the relationship $[\boldsymbol{a_{out}}] = [\boldsymbol{H}][\boldsymbol{a_{in}^n}]$ as depicted in Fig. 3A, the transfer matrix $\boldsymbol{H}$ was determined by its elements $\boldsymbol{h_{m,n}}$, which represented the impulse response function of the unit-lens in the lens array, where $\boldsymbol{m}$ and $\boldsymbol{n}$ represent the numbers of the unit-lens within the lens array in the horizontal and vertical directions, respectively. Considering the demagnification ($\boldsymbol{D}$) of the object image by the unit-lens, the $\boldsymbol{h_{m,n}}$ of the lens array can be defined as $\mathrm{rect}(\boldsymbol{xD/p}) * \mathrm{comb}(\boldsymbol{x/p})$, where rect($\boldsymbol{xD/p}$) and comb($\boldsymbol{x/p}$) are the window function of the unit-lens and the modulation of the entire lens array composed of many identical unit-lenses, respectively. In this study, all $h_{m,n}$ were assumed to be identical in this study. The transfer function can be described as $\mathrm{sinc}(\boldsymbol{v_x p/D})\mathrm{comb}(\boldsymbol{v_x p})$ through a Fourier transform of the impulse response, where $v_x$ is the spatial frequency of the displayed images. Thus, the lens array imaging can increase the amount of the transferable spatial information by a factor of $\boldsymbol{D}$.

**Printing experiment:** The integral lithographic system was implemented by modifying the optical platform in a conventional PμSL system comprised of a DMD-based digital microdisplay with a 405-nm LED source, the delivery optics, an optical diffuser (Thorlabs, DG100X100-1500), and the lens array, as shown in Fig. 1A. Note that the initial conditions of $\boldsymbol{R}$ and $\boldsymbol{A_S}$ for the PuSL machine used in this work is ~ 50 μm and $2.56 \times 10^3$ mm$^2$, respectively. According to the above relationship of $\boldsymbol{A_S = k \cdot R^2}$, the constant $\boldsymbol{k}$ was calculated to be $1.04 \times 10^6$. When investigating the scalable integral lithography process, we employed two types of lens arrays with different focal lengths and larger overall sizes, which were denoted as Lens 1 (RPC Photonics, MLA-S1000-f5.5, $\boldsymbol{f}$ = 5.5 mm, $\boldsymbol{p}$ = 1 mm, effective $\boldsymbol{NA}$ of 0.14 in the photopolymer, and overall size 50 × 50 mm$^2$) and Lens 2 (Flexible Optical B.V., APO-P(GB)-P250-F0.57, $\boldsymbol{f}$ = 0.57 mm, $\boldsymbol{p}$ = 0.25 mm, effective $\boldsymbol{NA}$ of 0.33 in the photopolymer, and an overall size 25 × 25 mm$^2$). For Lens 1, the ideal $\boldsymbol{k_{eff}}$ and $\boldsymbol{R_{eff}}$ were computed to be $1.35 \times 10^8$ and 4.35 μm using the relationship of $\boldsymbol{k_{eff} = k/D^2}$ and $\boldsymbol{R} \times \boldsymbol{D}$, respectively, where $\boldsymbol{D}$ is 0.087. For Lens 2, since the Abbe diffraction-limited spot size ($1.22\lambda$ / $2\boldsymbol{NA}$ = 0.74 μm) was larger than the effective planar resolution ($\boldsymbol{R} \times \boldsymbol{D}$ = 0.42 μm, where $\boldsymbol{D}$ is 0.0084), $\boldsymbol{R_{eff}}$ was considered to be 0.74 μm. Thus, the ideal $\boldsymbol{k_{eff}}$ was calculated to be $4.67 \times 10^9$ by applying the relationship of $\boldsymbol{k_{eff} = A_S/R_{eff}}^2$. In this work, the microstructures were printed at an imaging distance $\boldsymbol{b}$ of 68.75 mm. The photocurable material consisted of 1,6-hexanediol diacrylate (HDDA, Sigma-Aldrich) with a 2% (w/w) phenylbis (2, 4, 6-trimethylbenzoyl) phosphine oxide (Irgacure 819, Sigma-Aldrich) initiator and a 1-phenylazo-2-naphthol (Sudan 1, Sigma-Aldrich) a UV-absorber, which were varied in concentration from 0.05-0.5% (w/w). Also, we used commercial 3D printing resins (IC142-Investment Resin, Colorado photopolymer solutions) in our implementation of the integral lithographic fabrication system.

**Illumination scheme:** Increasing the illumination distribution over the lens array is an important factor in achieving scalable photopolymerization. Integral imaging is particularly beneficial for large-area printing because multiple superimposed array objects increase the area of uniform illuminance, as compared to the smaller region illuminated by a single object. In the proposed configuration, a digital microdisplay device projects dynamic images onto an optical diffuser, which functions as a virtual and reconfigurable photomask. The diffuser then scatters the light to produce a near Lambertian profile, which ensures homogeneous illumination in all directions at the plane of the lens array plane [25, 27, 28]. The scattered light enters the lens array that is positioned at an imaging distance of $\boldsymbol{b}$ in which each lens refocuses the light to create a smaller image of the images generated by the optical diffuser. The illumination distribution incident to the lens array was investigated using various object image configurations (see Figs. S8-10 in Supplementary Information). For simplicity, we used a circular shape as the virtual input image and assumed that the optical diffuser was an imperfect Lambertian emitter [50, 51]. This simplification allowed us to employ an adapted form of radiometric analysis [51, 52] when comparing the illumination distribution of a single object and that of an array of objects. From these assumptions, we derived approximated equations of illumination distributions for both a single object and array objects via a radiometric analysis using Cartesian coordinates (the details are provided in Supplementary Information). The calculated and measured illumination distribution in our imaging system is shown in Fig. S9. The illumination distribution was measured without the lens array by using a home-built scanner (XY-axis stepping motors) that included an optical powermeter and sensor (Thorlabs, PM100D and S120VC). To reproduce an illumination environment in which the light is incident immediately below the lens array, the optical power distribution was measured over an overall area of 50 × 50 $mm^2$ and a step size of 0.5 mm at the imaging distance $\boldsymbol{b}$ of 68.75 mm from the projected images (the details are provided in Supplementary Information). The measured results were plotted in the form of a 2D illumination distribution using MATLAB. The illumination distribution of a single circular source exhibited a narrow flat region, which provides limited options for scalability. However, the illumination homogeneity was significantly improved by superimposing array object sources. For example, the sum of the illumination distributions for a square array of 5 × 3 circular sources is depicted in Figs. S8D-F, and was uniform along the horizontal direction at an imaging distance $\boldsymbol{b}$ from the diffuser. Thus, these results indicate that this illumination superposition scheme in cooperation with the integral imaging method can be used to generates a large scale uniform illumination distribution.

**Acknowledgments**

S.K. and N.X.F. acknowledge support from Multidisciplinary University Research Initiative from the Office of Naval Research for financial support through Grant No. N00014-13-1-0631. N.X.F. acknowledges support by the U. S. Army Research Office through the Institute for Soldier Nanotechnologies at MIT, under Contract Number W911NF-13-D-0001. Y.T.C. acknowledges the National Research Foundation of Korea (NRF) grant funded by the Korea government (MSIT) (NRF-2017R1A2B4008053) and the Ministry of Trade, Industry & Energy (MOTIE, Korea) under Industrial Technology Innovation Program (No.20000665). J.J.H. acknowledges the mentorship and seed grant from the MIT Sandbox Innovation Fund. S.K. acknowledges Zheng Jie Tan and Kyungmin Sung for technical support of the illumination distribution measurement. S.K. acknowledges Jungki Song for proofreading.

**Author contributions**

S.K. and N.X.F. conceived the idea and directed the research. S.K., J.J.H., and Y.T.C. designed the experiments. S.K. developed the printing system and carried out the experiments. S.K., G.B., and N.X.F. analyzed and interpreted the results. N.X.F. supervised the whole project. S.K., J.J.H., and N.X.F. drafted the manuscript and all authors contributed to the writing of the manuscript.

**Competing interests**

N.X.F., J.J.H., and S.K. are inventors on an invention disclosure at MIT (case no. 20598; created May 31, 2018) related to this work. All authors declare that they have no other competing interests.

Supplementary information for

# Scalable additive manufacturing via integral image formation

Seok Kim[1], Jordan Jerome Handler[2], Young Tae Cho[3], George Barbastathis[1,4], and Nicholas Xuanlai Fang[1]*

[1]Department of Mechanical Engineering, Massachusetts Institute of Technology, Cambridge, MA 02139, USA.
[2]Sloan School of Management, Massachusetts Institute of Technology, Cambridge, MA 02142 USA.
[3]Department of Mechanical Engineering, Changwon National University, Changwon, South Korea.
[4]Singapore-MIT Alliance for Research and Technology (SMART) Centre, 1 Create Way, Singapore 138602.
*e-mail: nicfang@mit.edu

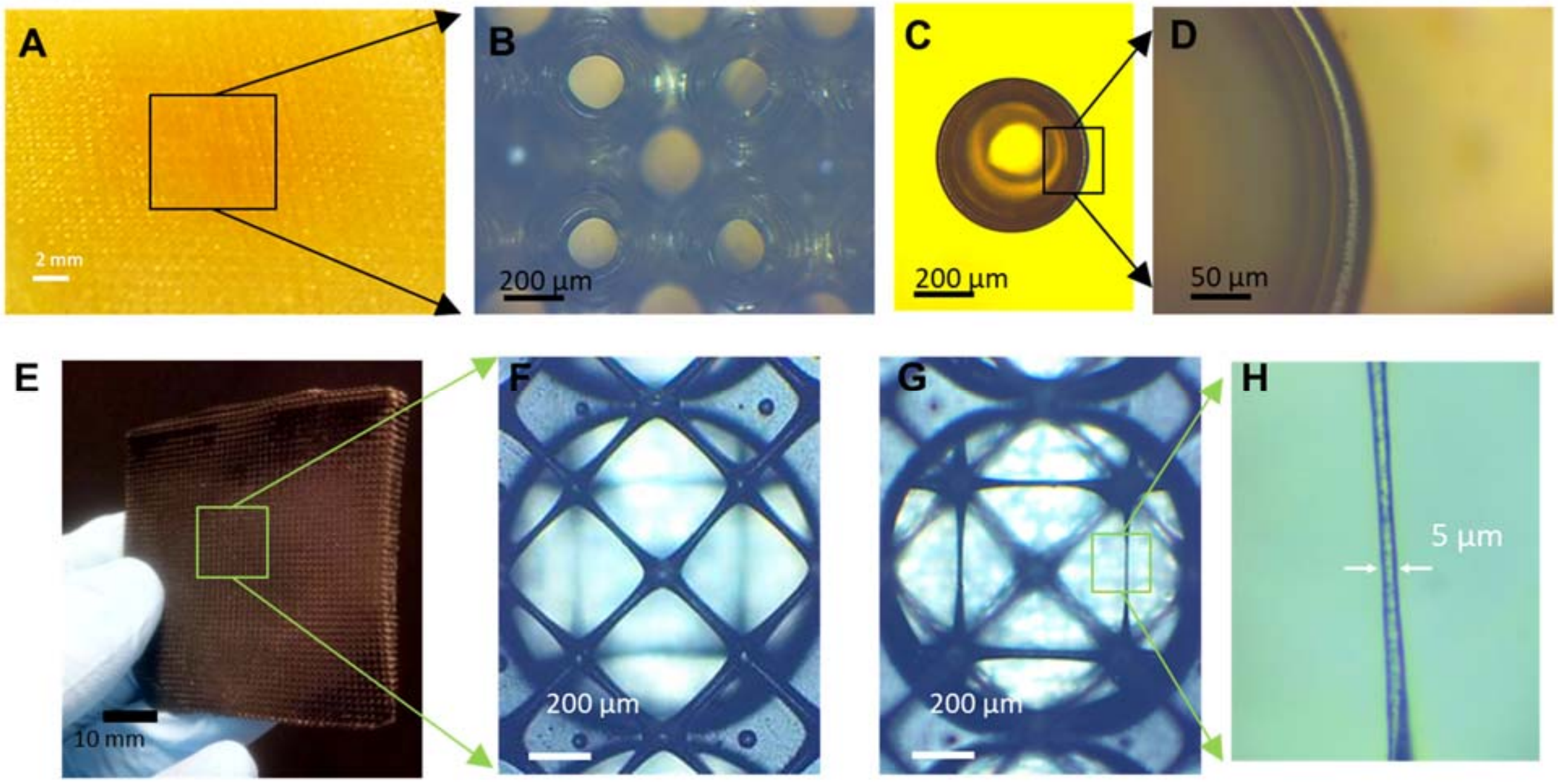

**Figure S1.** Optical microscope images of 3D-printed microstructures of Fig. 1 in the main text: (A-B) Triply periodic bicontinuous microstructures (total layers with slicing thickness of 50 µm: 60). (C-D) Trapezoidal shell-type microstructures with reentrant geometry (total layers with slicing thickness of 20 µm: 20). (E-H) An architecture of free-standing microfiber arrays with diameters 5 - 20 µm and an overall size of ~ 50 × 50 $mm^2$. (I) Cubic-truss microlattices (total layers with slicing thickness of 50 µm: 400).

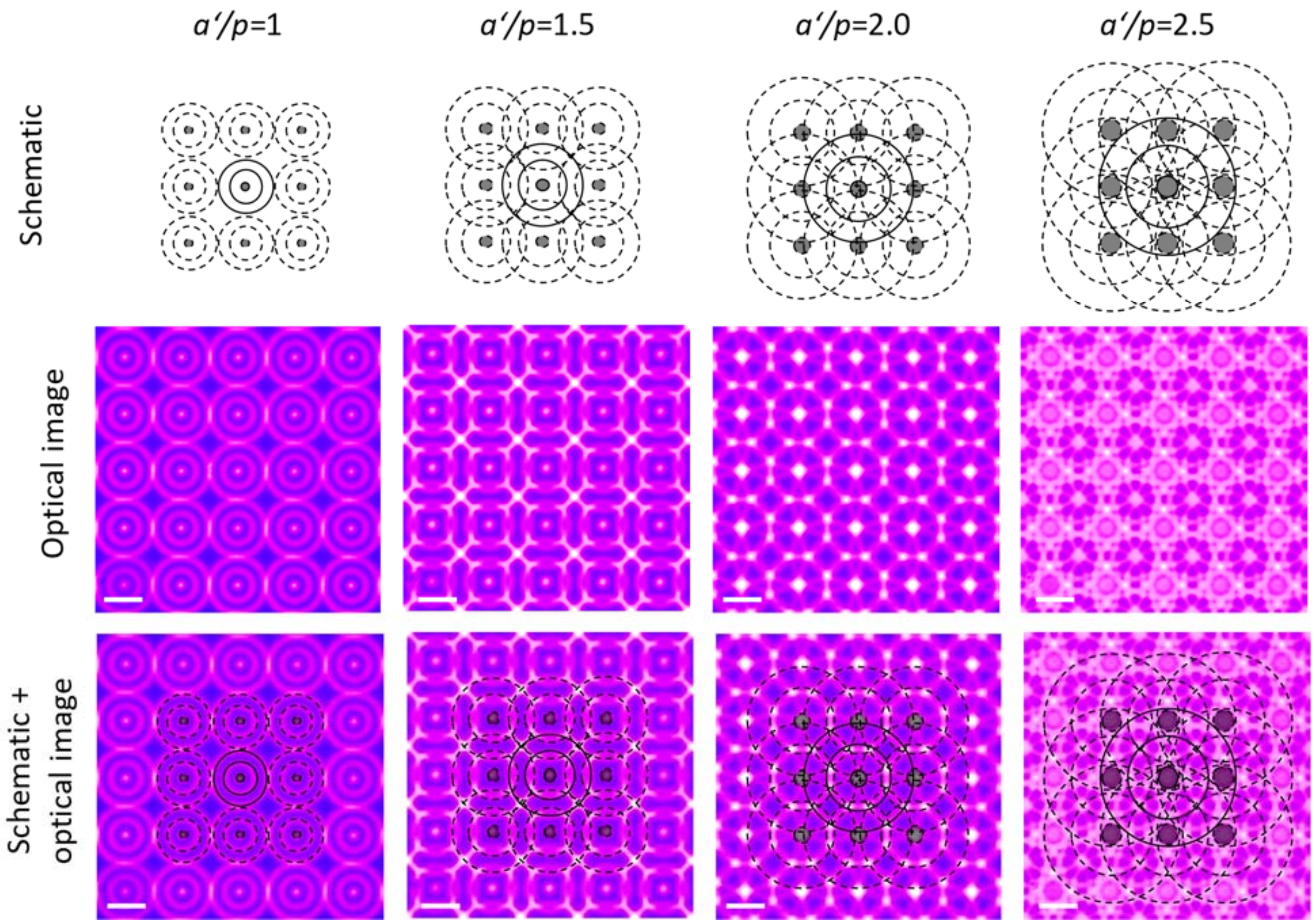

**Figure S2.** Image-based kaleidoscopic patterns with periodic optical distribution with varying ***a′/p***. Illustrative schematics and optical image are provided on the top and bottom, respectively. All scale bars are 100 μm.

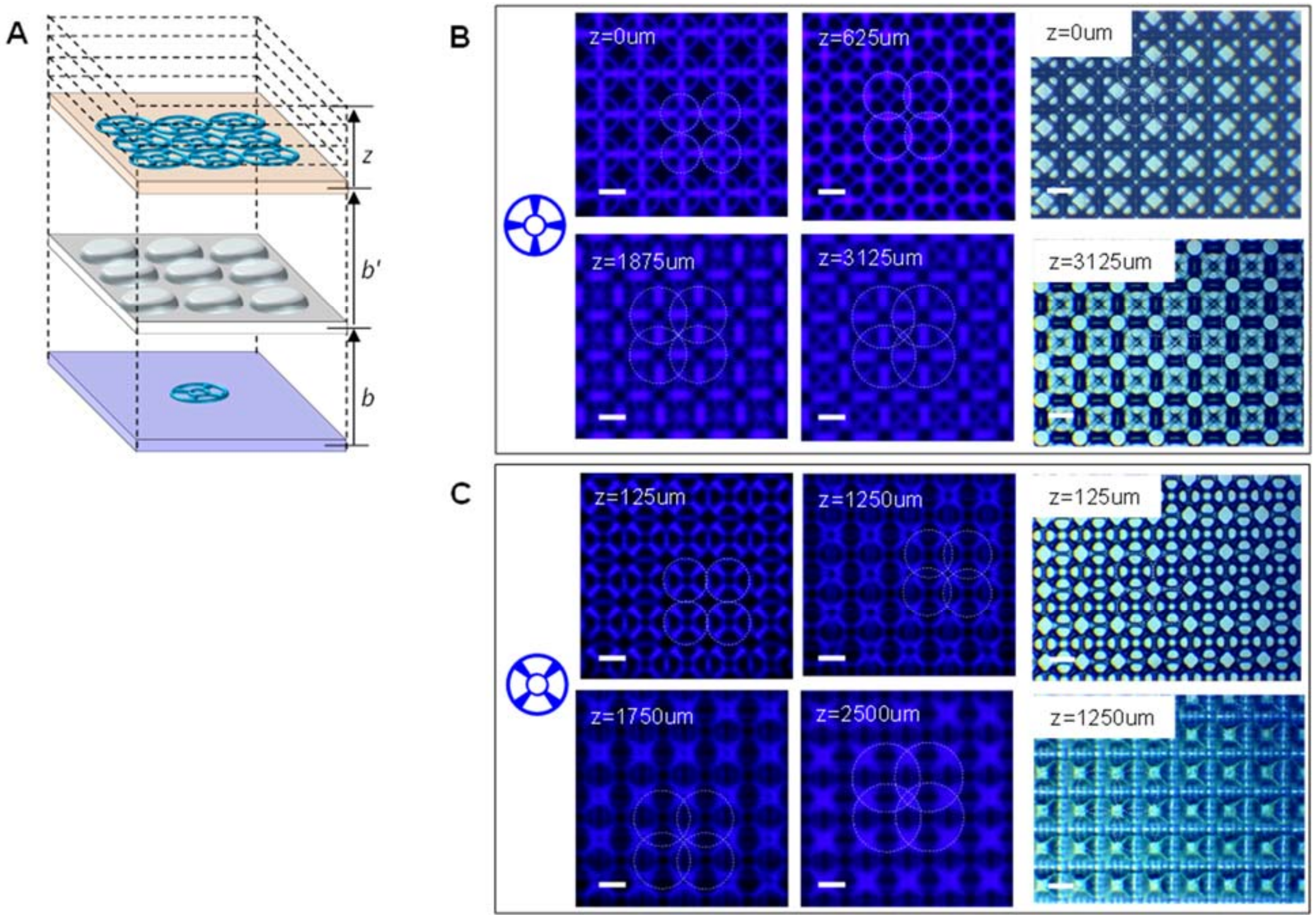

**Figure S3.** Out-of-plane image-based kaleidoscopic patterns with periodic optical distribution at different propagation distance along *z*. (A) Replicated patterns with an array-lens, including distance definitions. (B-C) Various kaleidoscopic images and printed structures. Object image is provided on the left. All scale bars are 100 μm.

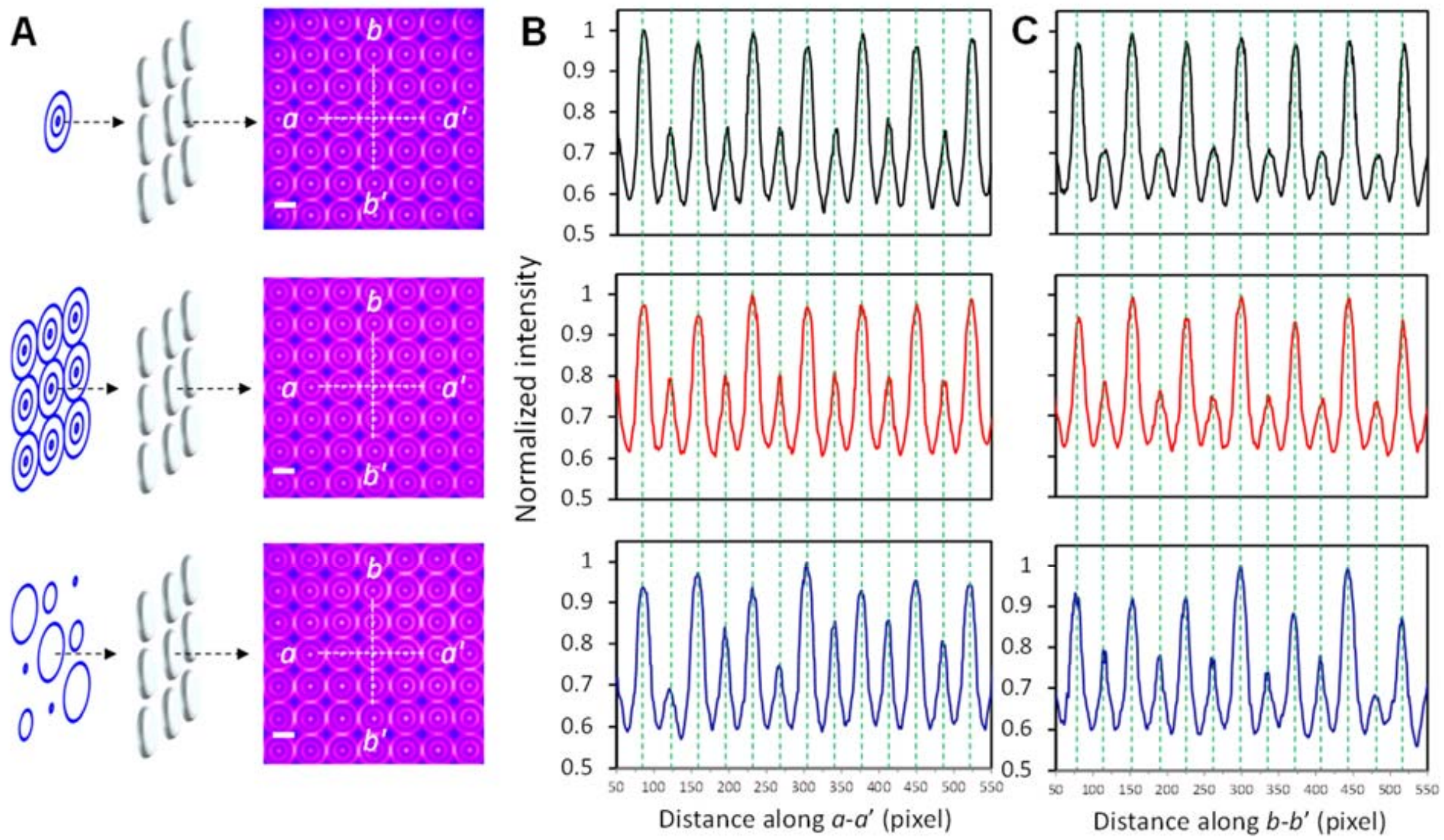

**Figure S4**. Intensity profiles of integral imaging patterns with compressive multi-projection. (A) Different imaging configurations and optical microscope images. Plot of normalized gray value versus distance along (B) line *a*-*a*' and (C) line *b*-*b*' in (A). A close analysis of captured images and their corresponding intensity profiles reveals that the synthetic pattern with three-decomposed EIs generates slightly asymmetric profiles as compared to that of synthetic patterns with identical EIs. This is presumably due to the small discrepancy of reconstructed images generated by the overlap of spatially non-uniform images from slightly different perspectives. All scale bars are 100 μm.

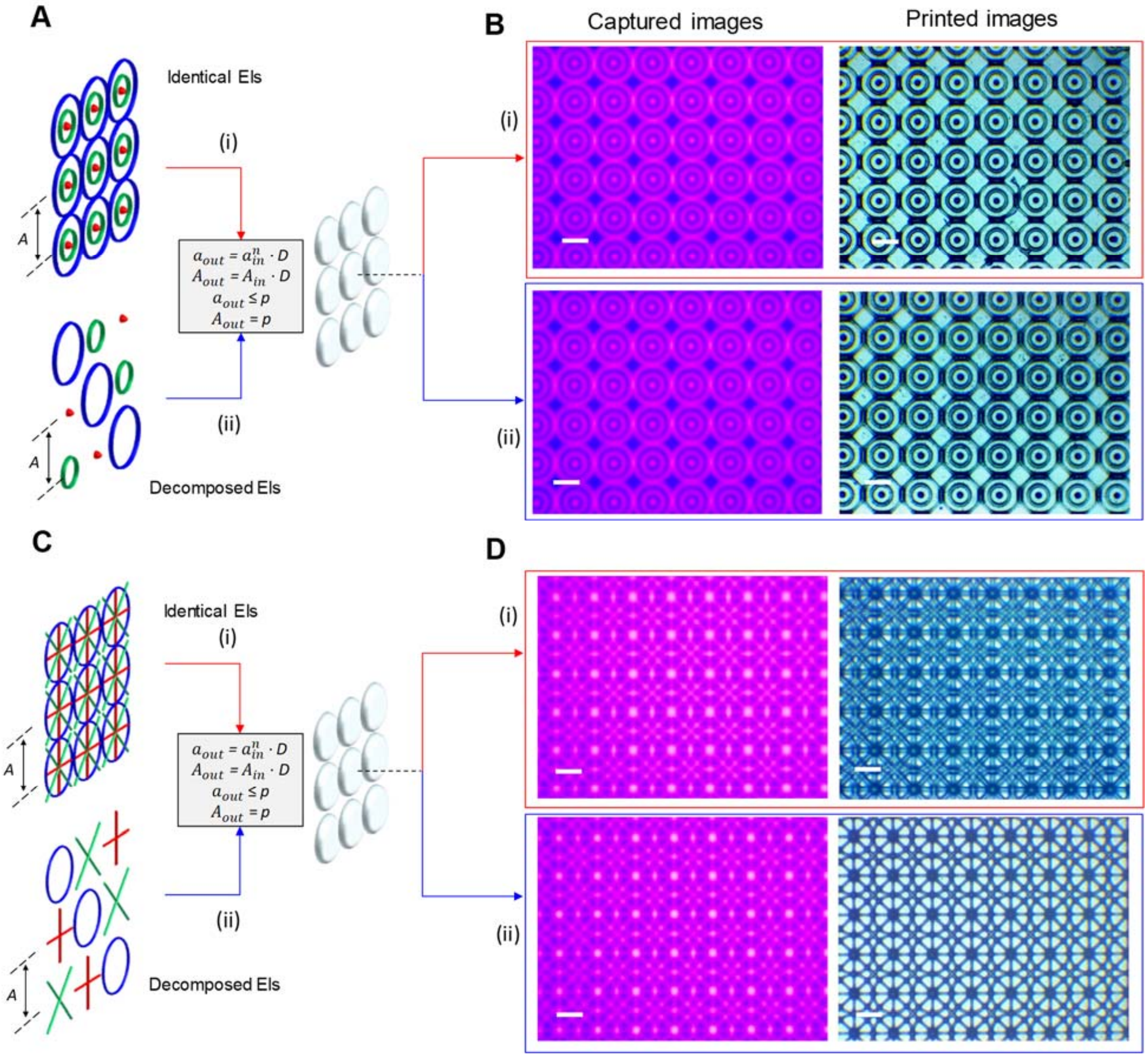

**Figure S5**. (A, C) Integral imaging mechanism by image replication and multiple EIs capture through the lens array. (B, D) Optical microscope-captured images and corresponding polymerized microstructures. All scale bars are 100 μm.

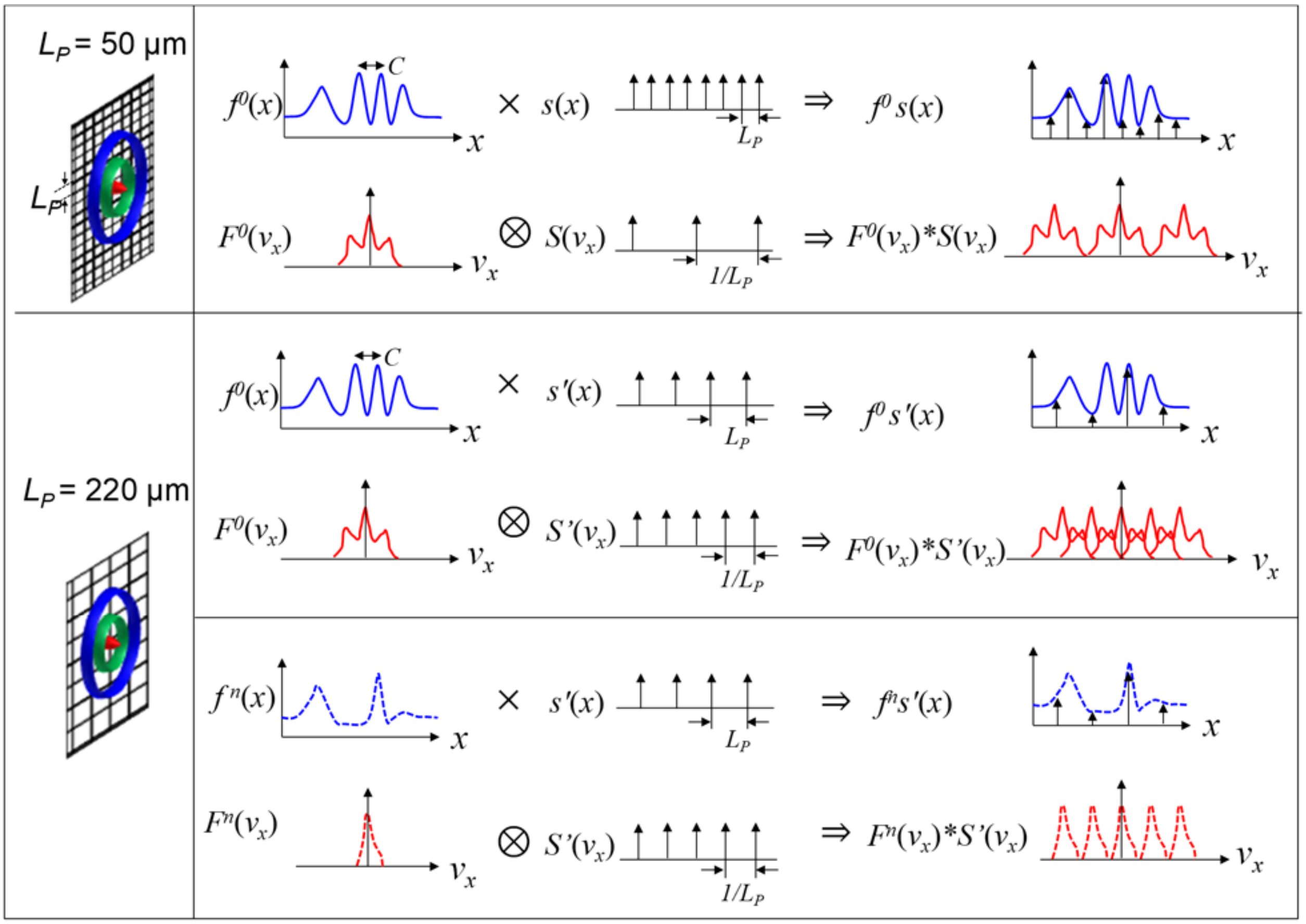

**Figure S6**. Frequency analysis of the image reconstruction for large (1280 × 800 with $\boldsymbol{L_P}$ of 50 μm) and low display bandwidth (1024 × 600 with $\boldsymbol{L_P}$ of 220 μm). Here, $\boldsymbol{f(x)}$ and $\boldsymbol{L_P}$ represent the target object and the projected pixel size of constituent digital microdisplay, respectively. $\boldsymbol{s(x)}$ and $\boldsymbol{v_x}$ denote the sampling function of comb($\boldsymbol{x/L_P}$) caused by the effect of the pixel sampling and the spatial frequency (= $1/\boldsymbol{C}$) of the target image, respectively.

| Reference | Type of digital micridisplay | Display area ($mm^2$) | Optics | Minimum feature size (μm) | Maximum Projection Area/exposure ($mm^2$) |
|---|---|---|---|---|---|
| 1 | DMD 1280 x 1024<br>pixel: 14–17 um | 17.9 x 14.3<br>21.76 x 17.4 | 5:1 reduction lens | 3.4 | 3.58 x 2.86<br>4.3 x 3.5 |
| 2 | DMD 1024 x 768<br>pixel: 13.68 um | 14 x 10.5 | N/A | 50 | N/A |
| 3 | LCD<br>pixel: 26 x 24 $um^2$ | N/A | N/A | 20 | 15 x 11 |
| 4 | DMD 1024 x 768<br>pixel: 13.68 um | 14 x 10.5 | NA 0.3<br>NA 0.13 | 10~30 | 1.95 × 1.95 |
| 5 | DMD 1024 x 768<br>pixel: 13.68 um | 14 x 10.5 | 10:1 reduction lens | 5 | 1.4 x 1.05 |
| 6-a,b,c | LCoS 1920 × 1080<br>pixel: 8 um | 15.36 x 8.64 | 6:1 reduction lens | 10 | 2.56 x 1.44 |
| 7 | DMD 1024 x 768<br>pixel: 13.68 um | 14 x 10.5 | 1:1 lens | 25~50 | 14.6 × 10.9 |
| 8 | DMD 1024 x 768<br>pixel: 13.68 um | 14 x 10.5 | 5:1 reduction lens | 2.5 | 2.8 x 2.1 |
| 9 | DMD 1024 x 768 | N/A | Built in lens of projector (CASIO XJ-S36) | 300 | 48 x 36 |
| 10 | DMD 1920 x 1080<br>pixel: 10.8 um | N/A | 2:1 reduction lens | 5.4 | 6 x 8 |
| 11-a,b,c | DMD 1920 × 1080 | N/A | N/A | 5 | 4.6 x 3.5 |
| 12 | DMD 1024 × 768<br>pixel: 13.6 um | 14 x 10.5 | 1:1 projection lens | 14 | 14 x 10.5 |
| 13 | DMD 1920 × 1080<br>pixel: 5 um | 9.6 × 5.4 | 1:4 projection lens | 200 | 384 x 216 |
| 14 | DMD 1024 x 768<br>pixel: 10.8 um | 11 x 8.3 | 5:1 reduction lens | 5 | 2.52 × 1.41 |
| 15 | LCoS 1920 × 1080 | | 7:1 reduction lens | 10~15 | 1.2 × 2.2 |
| 16 | DLP 1920×1080 | 20 | N/A | 100<br>75<br>50 | 190 × 110<br>142 x 78.7<br>94 x 53.3 |
| 17 | DLP 1920 x 1080 | N/A | N/A | vida: 73<br>vida HD: 50<br>Vida HD Crown & Bridge: 35 | vida: 140 x 79<br>vida HD: 96 x 54<br>Vida HD Crown & Bridge: 90 x 50 |
| 18 | DLP | N/A | N/A | DB9 creator: 70<br>B9 core 530: 30<br>B9 core 550: 50 | DB9 creator: 102 × 78<br>B9 core 530: 57.6 x 32.4<br>B9 core 550: 96 x 54 |
| 19 | DLP | N/A | N/A | PICO 2-39: 39<br>PICO 2-50: 50<br>PICO 2 HD27: 27<br>PICO 2 HD37: 37 | PICO 2-39: 51x32<br>PICO 2-50: 64x40<br>PICO 2 HD27: 51.8x29<br>PICO 2 HD37: 71x40 |
| 20-a,b,c | DLP | 10.3 x 7.7 | N/A | M1: 75<br>Experiment: 50 | 141 x 79 |
| 21 | DMD 912 x 1140<br>pixel: 7.6 um | 0.45" | 1:7 projecgtion lens | 50 | 64 x 40 |
| 22 | DMD 1024x768<br>pixel: 13.6 um | 14 x 10.5 | 10:1 reduction lens | 20 | 1.4 x 1.05 |
| 23 | Asiga Pico Plus 27<br>DMD 912 x 1140<br>pixel: 7.6 um | 0.45" | N/A | 60 | 35 × 21.8 |
| 24 | DMD 608 x 684<br>pixel: 7.6 um | 0.3" | 1:1 projection lens | 100 | 4x2 |
| 25 | DMD<br>pixel: 13.65 um | N/A | 7:1 reduction lens | 6 | 1.1 x 1.8 |
| 26 | DMD | N/A | NA 0.30 | 2.5 | 0.4 x 0.4 |
| 27 | Epson PowerLite S5<br>LCD 800 x 600<br>pixel: 17 um | 13.6 x 10 | NA 0.30 | 3~5 | 1.36 x 1 |
| 28-a,b | LCoS 1400 x 1050<br>pixel: 10.65 um | N/A | 1.5:1 reduction lens | 7.1 | 10 x 7.5 |
| 29 | Optoma HD20 HD<br>DLP 4500 | N/A | N/A | Optoma HD20 HD: 29.5<br>DLP4500: 51.2 | Optoma HD20 HD: 56.7 × 35.4<br>DLP4500: 65.6 × 41 |
| 30 | DLP 1024 × 768<br>pixel: 10.8 um | N/A | 1:3 projection lens | 30 | 32 × 24 |
| 31 | DLP 1024 x 768<br>pixel: 13.7 um | 14 x 10.5 | 10:1 reduction lens | Theo. 1.37<br>Exp. 5.8 | 1.4 x 1.05 |
| 32-a,b | DLP | N/A | N/A | Miicraft+ 56<br>MiiCraft 50: 30<br>MiiCraft 80: 41.5<br>MiiCraft 100: 53<br>MiiCraft 125: 65<br>MiiCraft 150: 78 | Miicraft+: 43 × 27<br>MiiCraft 50: 57 x 32<br>MiiCraft 80: 80 x 45<br>MiiCraft 100: 102 x 57.5<br>MiiCraft 125: 125 x 70<br>MiiCraft 150: 150 x 84.4 |
| 33 | LCoS<br>Pixels: 0.3 x $10^4$ | N/A | Holographic imaging system | 100 | 10 x 10 |
| 34 | DLP<br>Pixels: 1280 x 800 | N/A | Axial imaging system | 300 | up to 55 mm |

**Table 1**. Exposure area and minimum feature size of reported projection-based SLA printers. Commercial system performance is based on the manufacturer's specifications.

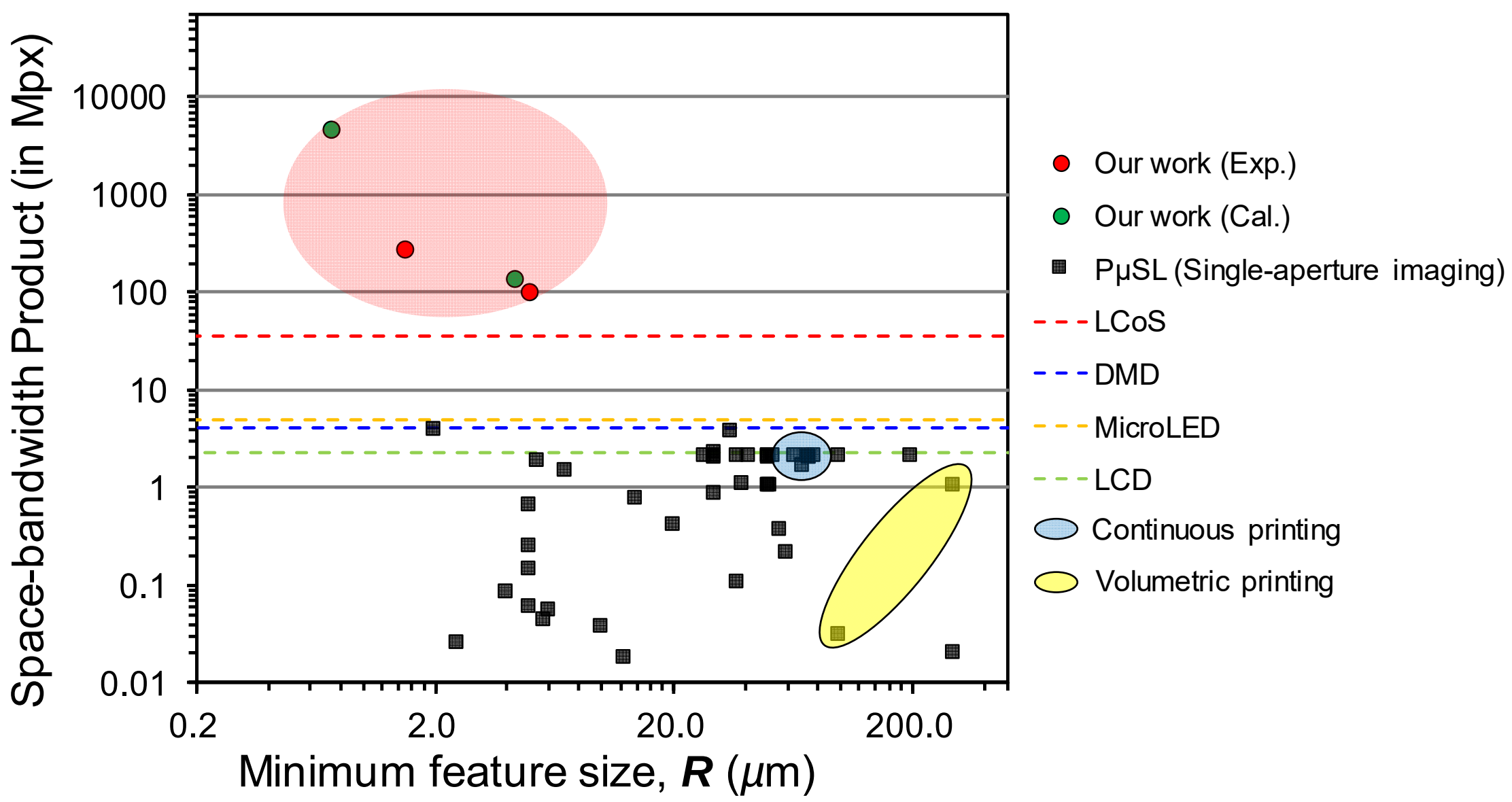

**Figure S7. SBP-resolution summary.** Plotted data points show the specific published results of the current PμSL systems. The dashed line represents analytical scaling equations grouped by digital microdisplay devices of LCoS, DMD, MicroLED, or LCD. The red and green circles represent experimental results as well as the authors' calculations for ideal results considering the potential of integral lithographic system, respectively.

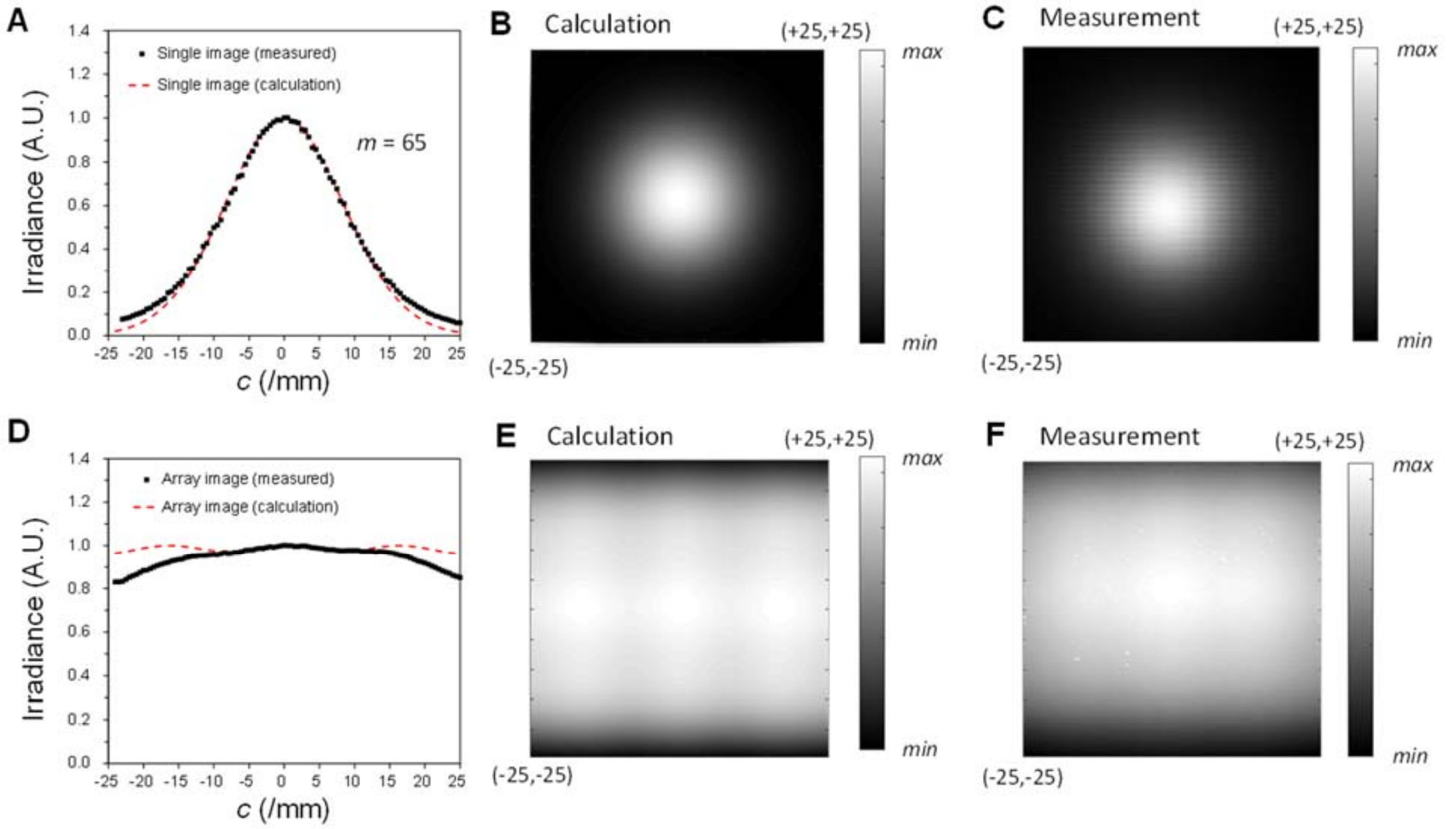

**Figure S8.** (A) The irradiance distribution (normalized to its maximum value) along the horizontal direction $\boldsymbol{C_x}$ direction at the center of vertical direction $\boldsymbol{C_y}$ for $\boldsymbol{a}$ = 9 mm and $\boldsymbol{b}$ = 68.75 mm. The dotted and dashed curves show the measured and calculated irradiance patterns of a single image source, respectively. (B-C) Calculated and measured 2D irradiance distributions for a single image source, respectively. (D) The corresponding irradiance distribution for a square array of 5 × 3 circular sources with $\boldsymbol{a}$ = 9 mm and $\boldsymbol{A}$ = 10 mm. (E-F) Calculated and measured 2D irradiance distributions for a rectangular array of image sources, respectively. We note that the irradiance was measured at the imaging distance $\boldsymbol{b}$ = 68.75 mm without the array-lens. All scale bars are 10 mm.

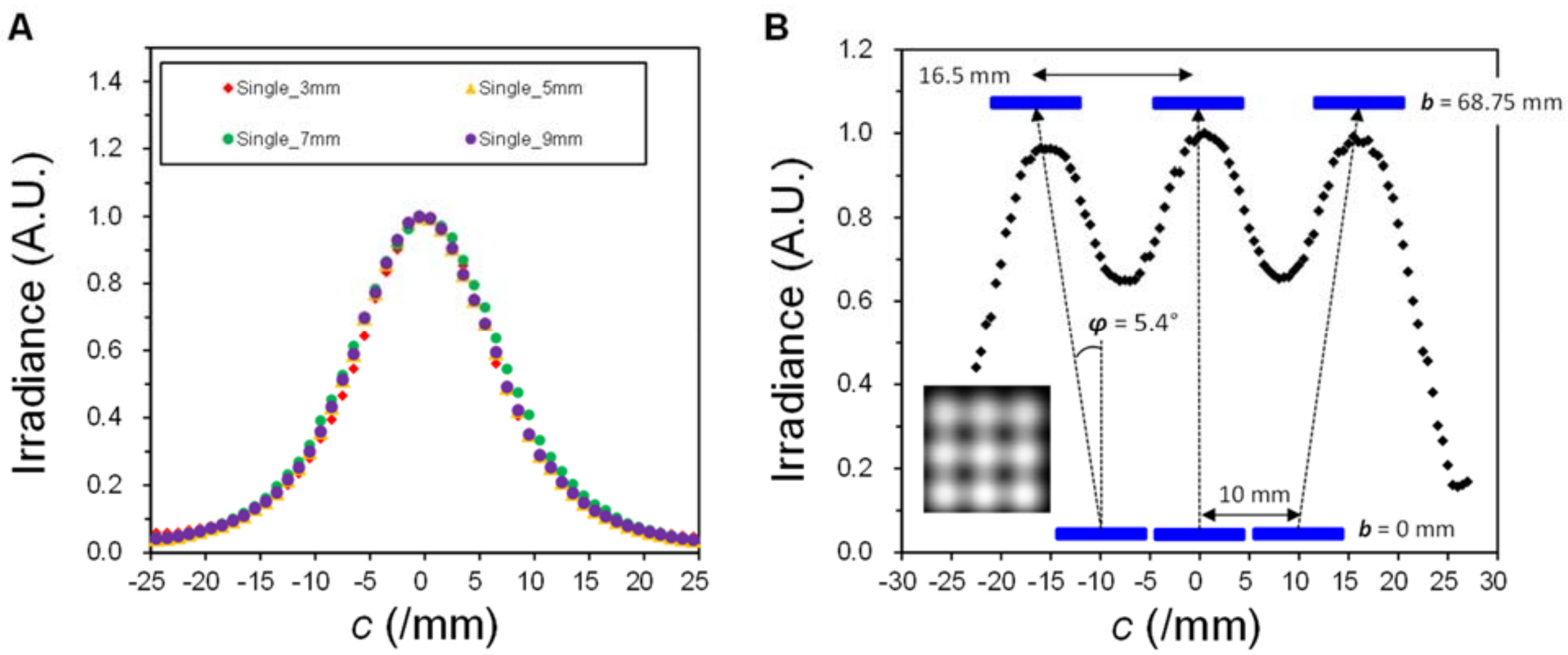

**Figure S9**. (A) The normalized illumination distribution according to the size variation of ***a***. (B) The measured divergence angle ***φ*** in our system. The irradiance distribution (normalized to its maximum value) was measured at the imaging distance ***b*** = 68.75 mm without the diffuser and the lens array.

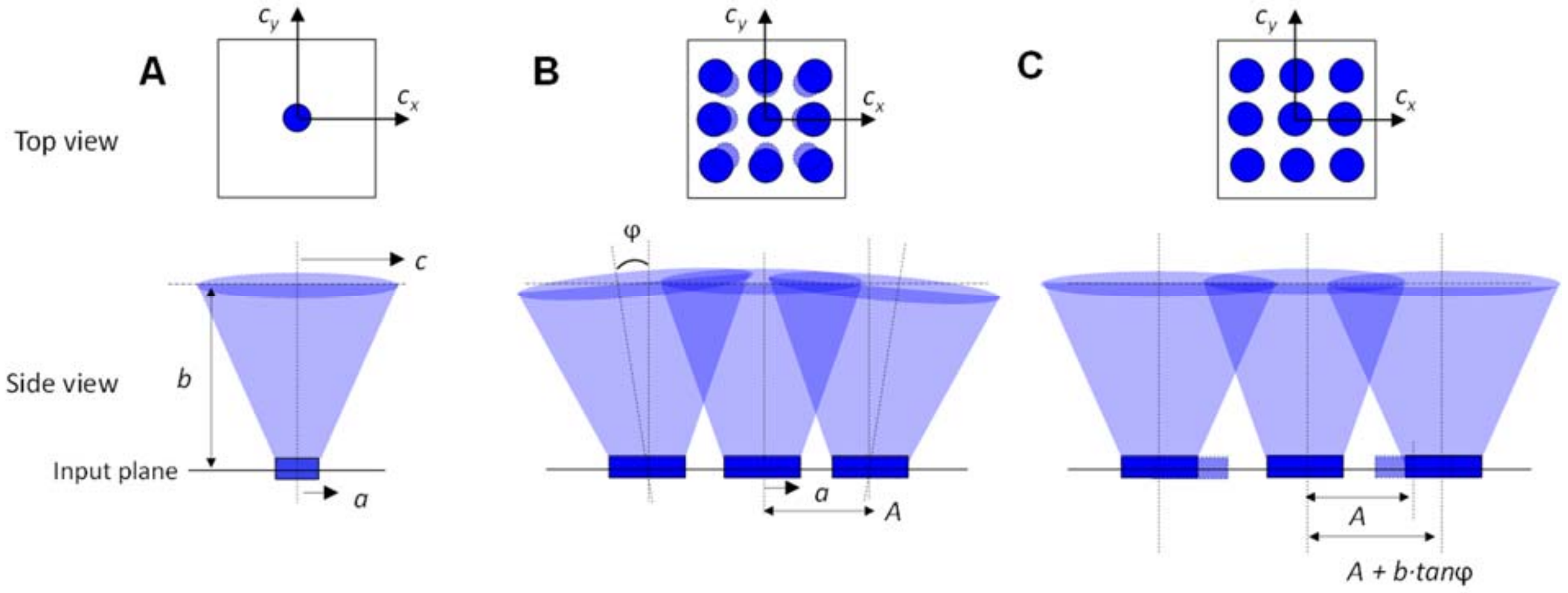

**Figure S10.** Conceptual schematics for analyzing the illumination distribution between a single image source and array image sources. (A) Illumination distribution of the single image source. (B) Actual illumination distribution and propagation of array image sources in our system. (C) The paraxial approximation for simplicity in this study.

**Illumination distribution**

To describe a simple analytical equation for the illumination distribution in our system, we assume that the input image produces an incoherent illuminating source and a nonperfect Lambertian distribution defined as a cosine-power function of $P(r,\theta) \propto P_0(r)cos^m\theta$ where $\boldsymbol{\theta}$ is the radiating angle, $\boldsymbol{m}$ is the directionality, $\boldsymbol{P_0(r)} = \boldsymbol{P_0/r^2}$ is the radiant power on the axis at distance $\boldsymbol{r}$ from the image source when $\boldsymbol{\theta} = 0°$, and $\boldsymbol{P_0}$ is proportional to $\boldsymbol{a}^2$. In Fig. S8A-C, the radiant power distribution projected at lateral position ($\boldsymbol{c_x}$, $\boldsymbol{c_y}$) over the illuminating plane at imaging distance $\boldsymbol{b}$ from a single image source is described by radiometric theory with Cartesian coordinates[48, 49]. Accordingly, the practical approximation for the radiant power distribution of a single image source right below the array-lens can be described as

$$P(a,b,c) \propto \frac{a^2 b^m}{[(b^2+c_x^2+c_y^2)]^{\frac{m+2}{2}}}. \quad \text{(S1)}$$

We measured the illumination distribution over the $\boldsymbol{c_x}$-$\boldsymbol{c_y}$ plane with variations of size $\boldsymbol{a}$ when $\boldsymbol{b}$ = 68.75 mm, and further fitted the measured irradiance distribution to determine the constant $\boldsymbol{m}$. Based on the measurements of the radiant power distribution for a single image source along the $\boldsymbol{c_x}$ axis at $\boldsymbol{c_y} = 0$ for a selected value of $\boldsymbol{a}$ = 9 mm, we extracted the fitting constant $\boldsymbol{m}$ = 65 in our system (Fig. S8A and 9A). Also, we noticed that the constant $\boldsymbol{m}$ significantly depends on the radiating characteristic of the diffusing optics and the divergence of the delivering optics, rendering it effectively independent of the size $\boldsymbol{a}$ (Fig. S9A). Consequently, we assume that all input images have equal values of $\boldsymbol{m}$ in this study. From the illumination distribution analysis with a single object source, we derive the approximated equation of irradiance superposition to estimate the illumination distribution from an array object source. After considering the measured divergence angle $\boldsymbol{\varphi}$ caused by the delivery optics of the initial printing system used in this study (Fig. S9B), the illumination distribution from the array object source is approximated by the summation of the irradiances of a matrix of $\boldsymbol{N} \times \boldsymbol{M}$ (where $\boldsymbol{N}$ and $\boldsymbol{M}$ are odd numbers) images,

$$P(a,b,c) \propto a^2 b^m \cdot \sum_{j=1}^{M} \sum_{i=1}^{N} \left\{ b^2 + \left[ c_x - (N+1-2i)\left(\frac{A+b\cdot tan\varphi}{2}\right) \right]^2 + \left[ c_y - (M+1-2i)\left(\frac{A+b\cdot tan\varphi}{2}\right) \right]^2 \right\}^{-\frac{m+2}{2}}. \quad \text{(S2)}$$

For the simple analytic approach, we used a paraxial approximation that the divergence angle shifts only the distance between array image sources and thereby model the illumination propagation to the array-lens without divergence (Fig. S10). In a simple circle array object source with $\boldsymbol{N} \times \boldsymbol{M}$ = 5 × 3, the calculated and measured illumination distributions for a rectangular arrangement of image sources is shown in Fig. S8E and F when $\boldsymbol{a}$ = 9 mm, $\boldsymbol{A}$ = 10 mm, and $\boldsymbol{\varphi}$ = 5.4 °. We believe that the discrepancy of the calculation and measurement at the edges may originate from the small divergence angle $\boldsymbol{\varphi}$ caused by the delivering optics in our initial printing system.

**References list of Table 1**